\documentclass[pre,aps,floatfix,superscriptaddress,twocolumn]{revtex4-1}
\usepackage{graphicx}
\usepackage{dcolumn}
\usepackage{latexsym}
\usepackage{hyperref}
\usepackage{amsmath, amsthm, amssymb}
\usepackage{epsfig}
\usepackage{bm}
\usepackage{geometry}
\usepackage[dvipsnames]{xcolor}

\begin{document}

\title{Adaptive dynamics of Ising spins in one dimension leveraging Reinforcement Learning}
\author{Anish Kumar}
\email[]{anishkumar.rs.phy22@itbhu.ac.in}
\affiliation{Department of Physics, Indian Institute of Technology(BHU), 
Varanasi- 221005, India}

\author{Pawan Kumar Mishra}
\email[]{pawankumarmishra.rs.phy19@itbhu.ac.in}
\affiliation{Department of Physics, Indian Institute of Technology(BHU), 
Varanasi- 221005, India}

\author{Riya Singh}
\email[]{riyasingh.rs.phy24@itbhu.ac.in}
\affiliation{Department of Physics, Indian Institute of Technology(BHU), 
Varanasi- 221005, India}

\author{Shradha Mishra}
\email[]{smishra.phy@iitbhu.ac.in}
\affiliation{Department of Physics, Indian Institute of Technology(BHU), 
Varanasi- 221005, India}

\author{Debaprasad Giri}
\email[]{dgiri.app@iitbhu.ac.in}
\affiliation{Department of Physics, Indian Institute of Technology(BHU), 
Varanasi- 221005, India}

\begin{abstract}
    A one-dimensional flocking model using active Ising spins is studied, where the system evolves through the reinforcement learning approach \textit{via} defining state, action, and cost function for each spin. The orientation of spin with respect to its neighbouring spins defines its state. The state of spin is updated by altering its spin orientation in accordance with the $\varepsilon$-greedy algorithm (action) and selecting a finite step from a uniform distribution to update position. The $\varepsilon$ parameter is analogous to the thermal noise in the system. The cost function addresses cohesion among the spins. By exploring the system in the plane of the self-propulsion speed and $\varepsilon$ parameter, four distinct phases are found: disorder, flocking, flipping, and oscillatory. In the flipping phase, a condensed flock reverses its direction of motion stochastically. The mean reversal time $\langle T \rangle $ exponentially decays with $\varepsilon$. A new phase, an oscillatory phase, is also found, which is a chaotic phase with a positive Lyapunov exponent.
    The findings obtained from the reinforcement learning approach for the active Ising model system exhibit similarities with the outcomes of other conventional techniques, even without defining any explicit interaction among the spins.
\end{abstract}

\maketitle
\section{Introduction}
Many-particle systems are fundamental in physics, engineering, and computational science, with applications ranging from microscopic to macroscopic scales. A many-particle system consists of a large number of particles, and at the individual level, particles can undergo complex interactions due to their surroundings. So, each particle has to adapt its dynamics according to its environment, which results in the collective behaviour of the system. The study of many-particle systems with conventional computer simulation approaches faces a number of challenges to a large number of degrees of freedom and the complex nature of many-body interactions \cite{cichos2020machine}. Even if we take care of interaction, implementing the true behaviour of interaction is not feasible. \\
Reinforcement Learning ($RL$) is a type of machine learning. It is a model-free approach and offers a powerful and flexible framework to address the complexity of interaction in many body systems.  $RL$ allows an agent to learn by its experience by taking action within the environment, and it receives feedback in the form of cost/reward \cite{dos2017adaptive,canese2021multi,hook2021learning, sutton2018reinforcement}. In $RL$, there are many model-free approaches like $Q$-learning \cite{watkins1992q,sutton2018reinforcement}, SARSA (State-Action-Reward-State-Action) \cite{mohan2021optimal}, Proximal Policy Optimization \cite{schulman2017proximal}, etc., where the agent directly learns the policy or value function through interaction with its environment. Such model-free methods only consider the observed states, actions, and rewards of the agent and pass over the stationary or non-stationary state of the environment.\\
In the present work, we focus on the use of $RL$ in active matter systems. Active matter systems are also many-particle systems, consisting of a large number of entities that take up energy from the environment and convert it into a directed motion. It is inherently far from equilibrium. It extends over a broad scale from bacterial colonies and cytoskeleton components to bird flocks, fish schools and Human crowds \cite{pincce2016disorder}. Janus particles \cite{walther2008janus,hu2012fabrication,zhang2017janus}, self-propelled colloids \cite{wang2015one,palacci2014light,buttinoni2013dynamical}, and swarming robots \cite{brambilla2013swarm} are some examples of non-living active matter systems. Due to their inherent nonequilibrium nature, these systems exhibit rich phenomena like collective behaviour and pattern formation. These characteristics make it very interesting for interdisciplinary areas. Rule-based models \cite{vicsek1995novel} have become very successful in predicting the basic features of active systems. One of the interesting characteristics of these systems is the collective coherent motion along a common direction, also called flocking \cite{vicsek2012collective,das2024flocking}.\\
Most of the flocking models are developed for two dimensions \cite{vicsek1995novel,chate2008collective}. But in general dimension of the system plays an important role in characterizing the properties of the flock.  Unlike higher dimensions, the one-dimensional flock shows a directional switching behaviour. They abruptly change their direction of motion in a cohesive manner, a characteristic useful for understanding how animal groups move.
In 1999, O`loan and M.R. Evans \cite{o1999alternating} proposed a one-dimensional lattice model for flocking in which particles can move either left or right. They found that the system goes from a homogeneous phase to a condensed (single flock) phase. This single flock reverses its direction of motion stochastically. There are more rule-based studies of one-dimension flocks that support the directional switching phenomenon \cite{czirok1999collective,raymond2006flocking,buhl2006disorder,yates2009inherent,bode2010making,dossetti2011cohesive, solon2013revisiting,sakaguchi2019flip,benvegnen2022flocking, KUMAR2024129773,pawanDirectionalCue, laighleis2018minimal}.\\
Recently, Durve {\em et al.} \cite{durve2020learning} showed that the cohesive nature of the flocking can be obtained from the minimization of the rate of neighbour loss using the $RL$ approach in two dimensions. 
Due to the self-propelled nature of active agents, the $RL$ approach is also useful for optimal navigation \cite{khlif2022reinforcement,pan2022research,gharbi2024dynamic}. In recent years, numerous studies have been done for the optimal navigation of active particles using the $RL$ approach in turbulent flows and under different potentials 
\cite{colabrese2017flow,nasiri2024smart,schneider2019optimal,putzke2023optimal,nasiri2023optimal,nasiri2022reinforcement,monderkamp2022active,biferale2019zermelo,alageshan2020machine,buzzicotti2020optimal,zou2022gait,falk2021learning,gerhard2021hunting,yang2020micro,muinos2021reinforcement}.\\
In this study,  we have developed the algorithm and explored the characteristics of a one-dimensional active Ising spin system using the $Q$-learning method.
For each spin, its state is defined based on whether its current location is in majority or minority and by using the $\varepsilon$-greedy algorithm, it takes action (whether to flip its orientation or not). Here, $\varepsilon$ is analogous to the thermal noise in the system. Since the spins are active, a hopping step or the self-propulsion speed is chosen randomly from a uniform distribution to introduce stochasticity in the system.\\
 The key findings of this study are:
(i) While exploring the system in a plane of self-propulsion speed and $\varepsilon$ parameter, by observing the behaviour of the order parameter time series and distribution of the spins in the system, we have identified distinct phases, namely disorder, flocking, flipping and oscillatory phase.
(ii) The flipping phase is a phase where the flock stochastically reverses the direction of motion. The mean reversal time of the flock decays exponentially with the $\varepsilon$ parameter.
(iii) A new oscillatory phase is also found in which the system reverses its direction of motion at every time step.  Further, we find that the oscillatory phase is chaotic in nature with a continuous power spectrum and positive Lyapunov exponent.\newline

The article is organized as follows. In Sec. \ref{II}, we discuss the details of the model, $Q$-learning scheme, state and action choices. And then show the results in Sec. \ref{III}. In Sec. \ref{III A}, we explore the phase diagram of the system, followed by the discussion of the flipping phase in Sec. \ref{III B}. Finally, we present the conclusion in Sec. \ref{IV}.

\section{Model}\label{II}

\begin{figure*}
\centering
{\includegraphics[width=0.75\linewidth,height=0.35\textheight]{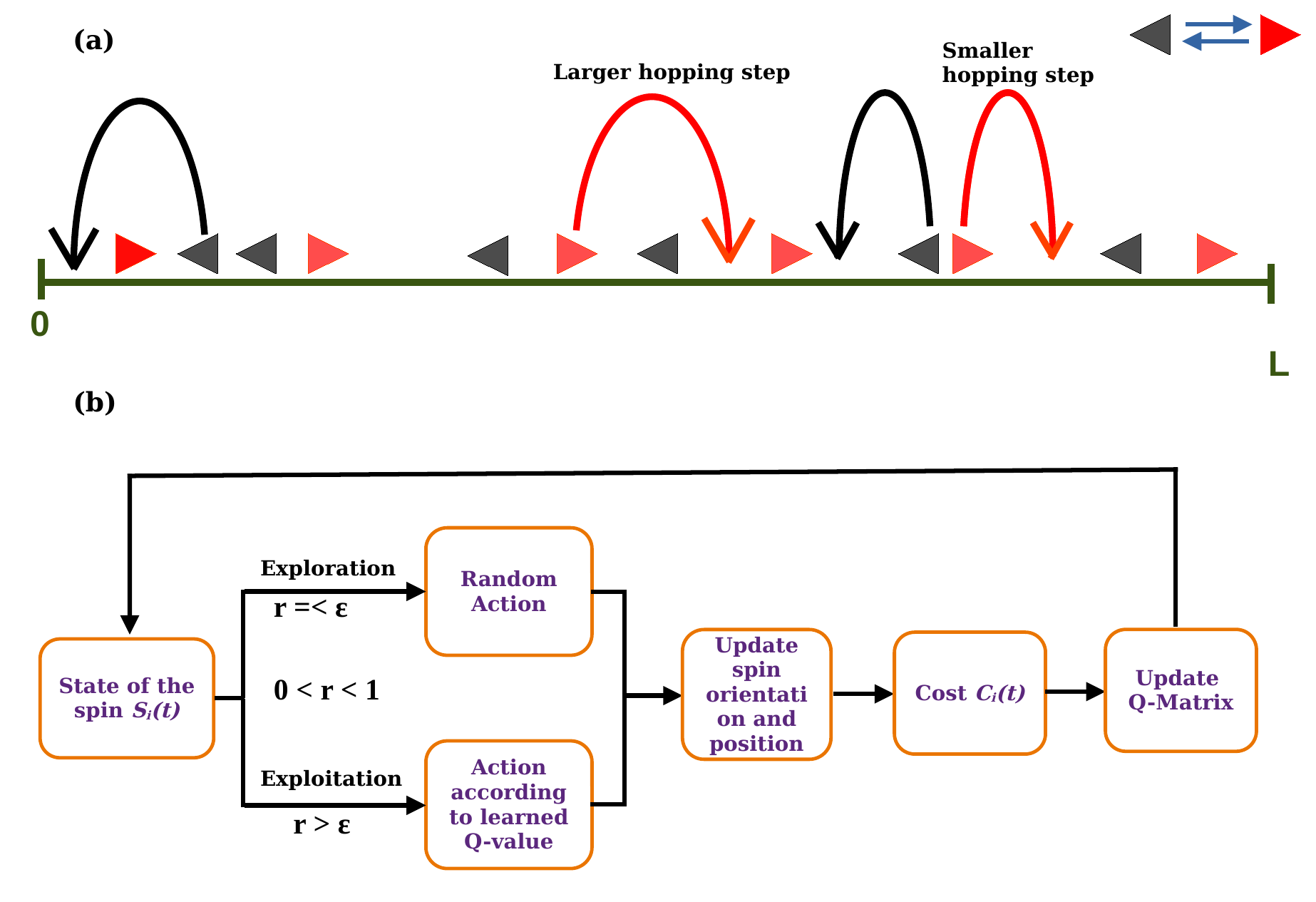}}
    \caption{(a) Schematic representation of one dimensional active Ising Model with spin up and down with solid triangles filled with red and black colors, respectively. The curved arrows show the hopping step of the spins and the direction of motion. The spin with up(down) orientation always moves towards the right(left).  Since $\Tilde{v_i}$ is taken randomly from a uniform distribution, the curved arrow has a different curvature. The spin can change its orientation based on the $\varepsilon$-greedy algorithm. (b) The Schematic diagram represents the $Q$-learning algorithm. The state of each spin is defined based on its majority or minority within its neighbourhood. The action for the spin is to flip its orientation or not. It takes action according to the $\varepsilon$-greedy algorithm, where the spin explores the system with $\varepsilon$ probability by taking random action. By updating spin orientation and changing the position, spin updates its state, and based on the updated state, the cost is decided. According to Eq.~\ref{matrix}, using the cost value, $Q$- matrix updates.}
\label{fig:1}
\end{figure*}

We start with $N$ active Ising spins $(s = \pm 1)$ that are randomly distributed along a length $L$ with periodic boundary conditions. To incorporate the reinforcement learning approach, we define the state and action for each $i^{th}$ spin. The state of each $i^{th}$ spin is determined based on the direction of the spin with respect to its neighbours within the range $[x_i - \delta x, x_i + \delta x]$, where $x_i$ is the position of $i^{th}$ spin and $\delta x$ is the interaction range and chosen as the unit of length in the system. In this case, there are two possibilities: the spin is aligned either in the same or opposite direction, with the majority of the spins within the range. So for each $i^{th}$ spin at any instantaneous time $t$, we have two states: $S_i(t) = \pm 1$. If the spin has the same direction as the majority of spins in the range, then its state $S_i(t) = +1$ and $S_i(t) = -1 $ for the opposite direction. Since the spin is Ising type, it has only two possible directions. Further, the action $a_i(t)$ for each spin is either to flip or keep its orientation.\\
Position update for the spin at each time step $\Delta t$ is according to:
\begin{equation}
    x_i(t+\Delta t) = x_i(t) + \Tilde{v_i}(t) s_{i}(t+\Delta t) \Delta t
    \label{position}
\end{equation}
here, $x_i(t)$ and $\Tilde{v_i}(t)$ are the position and instantaneous self-propulsion speed of the $i^{th}$ spin at time $t$, respectively, and $s_i(t + \Delta t)$ is the updated spin orientation of the $i^{th}$ spin. $\Tilde{v_i}(t)$ is taken from a uniform distribution with nonzero positive lower and upper bounds $v_1$ and $v_2$, respectively.  In most of the previous studies of active systems in $two$ and higher dimensions, the self-propulsion speed is constant for all the particles at all time steps \cite{vicsek1995novel,pattanayak2018collection,singh2021ordering,singh2021bond,czirok1999collective}. However, in $one$ dimension, the constant self-propulsion speed will lead to the motion of the particles on the lattice. To avoid it, we choose the self-propulsion speed from a distribution. Where the lower and upper limits of the distribution $v_1$ and $v_2$ decide the range of self-propulsion speed in the system. Additionally, the upper limit is varied from $v_1$ to $10$ times $v_1$.
At every time step, the magnitude of $\Tilde {v_i}(t)$ is chosen randomly; hence, at every instance, each particle can take a random step size obtained from the distribution. The inhomogeneous speed of individuals is also observed in experiments on fish school \cite{katz2011inferring} and bacterial colony \cite{cisneros2011dynamics}. The lower limit of the self-propulsion speed varied in the range $(10^{-5} - 10^{-1})$ to vary the activity in the system. The translation motion of the spin is purely due to activity and does not utilise the reinforcement learning approach explicitly.  \\
We mentioned the word {\em explicitly} because indirectly, even translational hopping involves learning. The measure of learning is adopted by maintaining the cohesion among the spins. So, the spin receives feedback when it moves to a new position. The spin pays a cost if it loses the number of neighbours around it. Hence the cost function $C_i(t + \Delta t)$ for each spin is defined as: 
\begin{equation}
C_i(t+\Delta t)=
    \begin{cases}
        1, & \text{if } n_i(t+\Delta t) < n_i(t)  \\
        0, & \text{ otherwise }  \\
    \end{cases}
    \label{cost}
\end{equation}
Here $n_i(t)$ number of spins within the range $[x_i - \delta x, x_i + \delta x]$ at time $t$. Further, for each $i^{th}$ spin, there is a $Q$ matrix made up of state-action pair for each spin. Since we have two states and two actions for each spin in this system so, for each $i^{th}$ spin, we have a $2$x$2$ dimensional $Q_i[S_i(t), a_i(t)]$ matrix. The $Q$ matrix is updated at every time step. We initialize it by having  all the entries of the matrix as zeros, and at each time, it is updated according to the following equation:
\begin{equation}
    Q_i[S_i(t), a_i(t)] \longleftarrow Q_i[S_i(t), a_i(t)](1 - \alpha) + \alpha C_i(t+\Delta t)
    \label{matrix}
\end{equation}
here, $\alpha$ is the learning rate. It is a parameter that controls the influence of the new information over the current $Q$ value. Both very high and low learning rates should be avoided for optimized learning. For a large learning rate, the system may skip some of the states of the phase space \cite{sampat2022ordering}, whereas very weak learning can result in slow convergence. \\
Further, the action is chosen based on $\varepsilon$-greedy algorithm \cite{dos2017adaptive,sutton2018reinforcement}. That is defined as:
\begin{equation}
a_i(t)=
\begin{cases}
    \text{random action}, & \text{with probability} \hspace*{0.3em}\varepsilon \\
    \text{argmin}\hspace*{0.3em}{Q_i[S_i(t),a_i(t)]}, & \text{with probability}\hspace*{0.3em}  (1- \varepsilon)\\
\end{cases}
\label{epsilon}
\end{equation}
so with $\varepsilon$ probability, spin takes random action (exploration), and with $(1-\varepsilon)$, it will choose an action such that for that state-action pair, $Q$ value is minimum (exploitation). Because the minimum $Q$ value tells that the cost value is also minimum for the corresponding state-action pair. 
So, the two parameters $\varepsilon$ and $\alpha$ control the system's exploration and exploitation. For the current study, we fix the learning rate $\alpha$ and vary the $\varepsilon$. The details of the model and algorithm have been explained in the Fig.~\ref{fig:1}(a) and (b), respectively. \\
To characterize the ordering in the system, we define average magnetization $\langle m \rangle$ as the order parameter. 
\begin{equation}
    \langle m \rangle = \frac{1}{N}\sum_i s_i\,
\end{equation}

The value of $|\langle m \rangle|$ lies in the interval $(0,1)$.
When all the spins have similar orientation, then  $|\langle m \rangle| \simeq 1$, and if all spins are randomly orientated, then $|\langle m \rangle| \sim 0$. \\
In the simulation, we have used $\alpha= 0.001$, $\delta x = 0.01$, $L = 100 \delta x$-$300 \delta x$, $\Delta t = 1.0$ and the mean density $\rho_0 = \frac{N}{L^2} = 20$.  The self-propulsion speed ${v_1}$ varies from  $1\times10^{-5}$ to $1\times10^{-1}$ . We define the dimensionless self-propulsion speed $\bar{v} = \frac{v_1 \Delta t}{\delta x}$. Hence $\bar{v} $ can vary between $10^{-3} $ to $10$. Since the speed of a particle in a single run is obtained by a distribution $(v_1, v_2)$, hence we also define mean self-propulsion speed $v_0 = \frac{(v_1 + v_2)\Delta t}{2 \delta x}$. The $ \varepsilon$ is varied between $[0.01, 1.0]$. The total simulation time step is $ 2 \times 10^5$. The thermal averaging is performed over the last $1.5 \times 10^{5}$ number of time steps. The system is studied for $N = 1000$, $3000$ number of Ising spins. For better statistics, the system is averaged over $100-200$ independent realizations.

\section{Results}\label{III}
Now we start analyzing the results of the system by varying the two parameters $\epsilon$ and $\bar{v}$. 
\subsection{Phase Diagram}\label{III A}
\begin{figure*} 
\centering
{\includegraphics[width=0.8\linewidth]{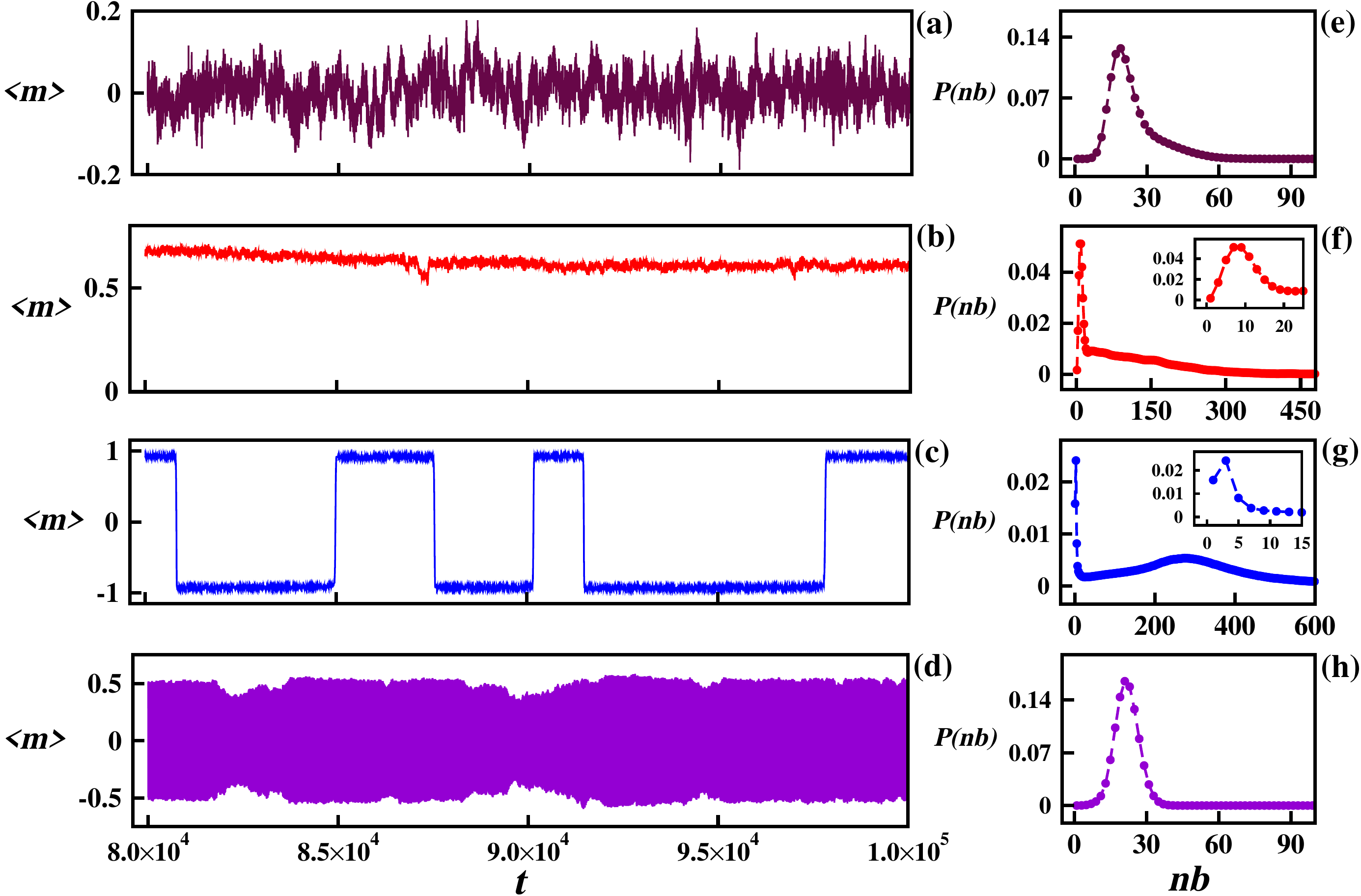}}
    \caption{(color online) Plot (a-d) illustrates the time series of order parameter $\langle m \rangle$ for different self-propulsion speed  $v_0 = 0.003, 0.03, 0.30, 3)$, respectively and $\varepsilon = 0.020$ in the steady state. And (e-h) shows the neighbour distribution $P(nb)$ of the spins in the system for the similar values of $v_0$, respectively.  Here, $N = 3000, L = 300 \delta x$.  
    }
\label{fig:2}
\end{figure*}
We started with initial random orientation and homogeneous distribution of the spin and let the system evolve to the steady state. The steady state is defined when the characteristics of the system remain statistically similar with respect to time. We first explore the ordering in the system with the variation of self-propulsion speed. In Fig.~\ref{fig:2}(a-d), we show the order parameter time series in the steady state for different $v_0$ = $(0.003, 0.03, 0.30, 3)$, respectively and for $\varepsilon = 0.020$. In Fig.~\ref{fig:2}(a), for very small self-propulsion speed $v_0 = 0.003$, there is no ordering in the system.  Hence $\langle m \rangle \approx 0$. In Fig.~\ref{fig:2}(b), when $v_0 = 0.03$, the system shows finite ordering where the order parameter fluctuates around its mean value  $\langle m \rangle \approx 0.60$, and ⟨m⟩ remains the same sign throughout all the depicted time steps. As the value of self-propulsion speed increases to $v_0 = 0.30$, as shown in Fig.~\ref{fig:2}(c), the system shows global ordering where the value of $\langle m \rangle$ is almost $\pm 1$. The order parameter $\langle m \rangle$ shows switching from positive to negative values, and vice versa. Further increase in speed increases the frequency of switching. In Fig.~\ref{fig:2}(d), we show the time series of $\langle m \rangle $, for $v_0 = 3$. Here, the value of order parameter $\langle m \rangle \approx \pm 0.5$, but $\langle m \rangle $ switches from positive to negative at every time step.\\

To further analyze the effect of self-propulsion speed on the steady-state configuration of the spins in the system, in Fig.~\ref{fig:2}(e-h), we have shown the probability distribution of the number of neighbours around the spin $P(nb)$  for the same values of self-propulsion speed and $\varepsilon$ as discussed in Fig.~\ref{fig:2}(a-d), respectively. To calculate the neighbour distribution $P(nb)$, we count the number of neighbours $(nb)$ within the interaction range $[x_i - \delta x, x_i + \delta x]$ for each spin and make the histogram, which is averaged over many snapshots in the steady state and different realizations. In Fig.~\ref{fig:2}(e),  the distribution $P(nb)$ is shown for $v_0 = 0.003$. The distribution shows a peak around $nb = 20$, which is the mean number density of the system, with a short tail. This shows that spins are moving in the system with a nearly homogeneous distribution. For $ v_0 = 0.03$, $P(nb)$ shows a peak around $nb \approx 10$ (as shown in the inset), and a long tail as shown in Fig.~\ref{fig:2}(f). The peak with the long tail in the distribution indicates a large dense cluster in the sea of small clusters of diluted density, $nb \approx 10$, i.e. the distribution of the spins in the system is inhomogeneous. As the self-propulsion speed further increased to $v_0 = 0.30$, $P(nb)$ shows bimodality as shown in Fig.~\ref{fig:2}(g). The first peak is at $nb \approx 3$ (as shown in the inset), and the second peak is at higher $nb  \approx 300$. This shows that there is a large dense cluster moving in the system with a few random moving spins, i.e. the system has a very inhomogeneous distribution of spins. In Fig.~\ref{fig:2}(h), the neighbour distribution $P(nb)$ is shown for $ v_0 = 3$, where again the distribution shows a single peak around the mean density of the system $nb = 20$, i.e. the spins configuration is again homogeneous in the system. \\

\begin{figure} 
\centering
{\includegraphics[width=0.92\linewidth,height=0.30\textheight]{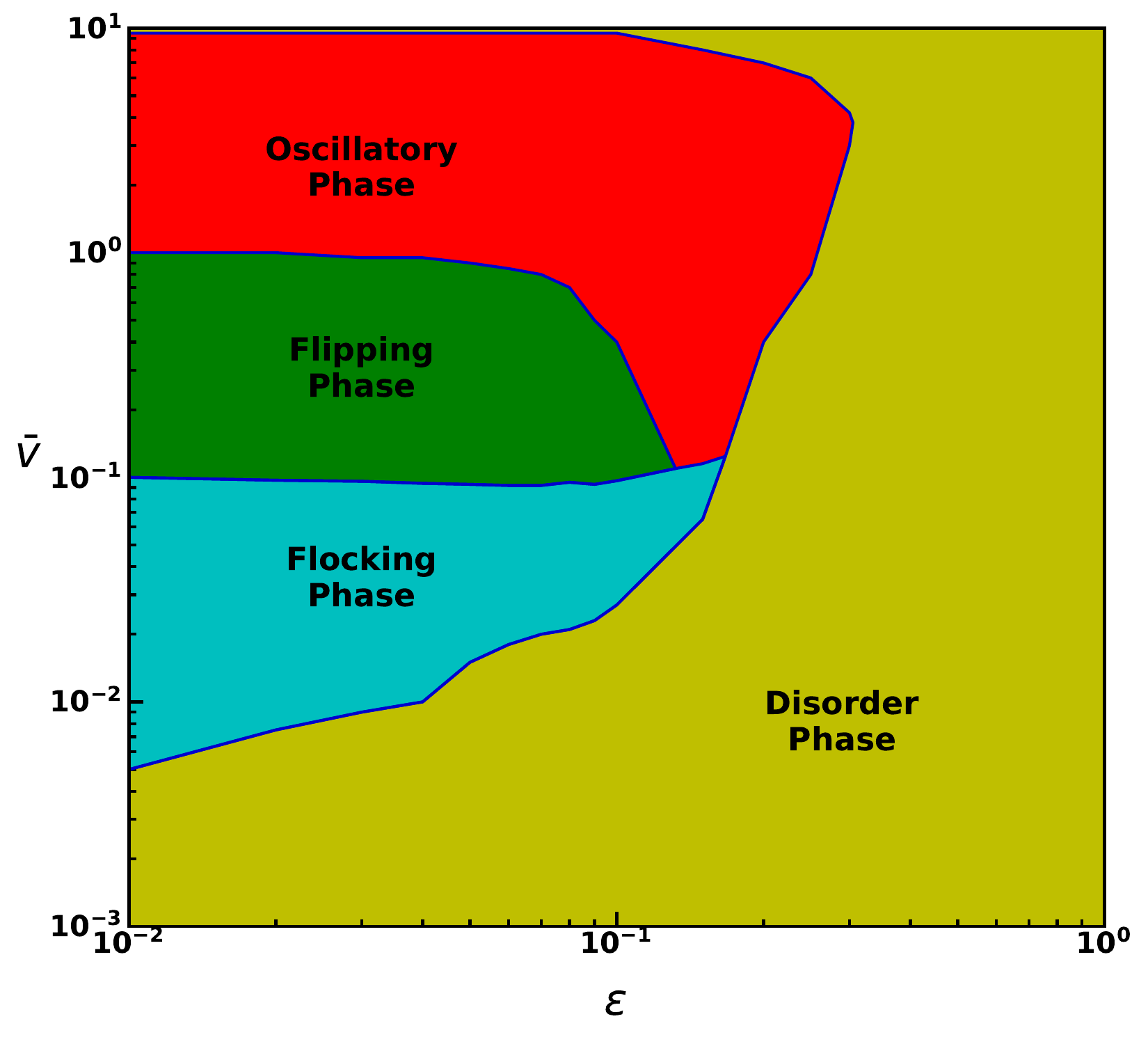}}
    \caption{(color online) This diagram illustrates the phase diagram of the system in the plane of $\varepsilon$ and $\bar{v}$, where different colors represent different phases. Here, $L = 300 \delta x$, and $N = 3000$.   }
\label{fig:3}
\end{figure}

By observing the behaviour of order parameter $\langle m \rangle$ and density distributions $P(nb)$ as shown in Fig.~\ref{fig:2}(a,e), (b,f), (c,g) and (d,h), respectively, we can identify the four distinct phases in the system named as (i) disorder, (ii) flocking, (iii) flipping, and (iv) oscillatory based on their characteristics. \\
In Fig.~\ref{fig:3}, we show the phase diagram of the system in the plane of self-propulsion speed $\bar{v}$ {\em vs.} $\varepsilon$.  \\

(i) {\bf{\textit{Disorder Phase}}}: For very low self-propulsion speed   $\bar{v} \sim 0.001$ and a wide range of $\varepsilon$, the yellow-shaded region in Fig.~\ref{fig:3} shows the disorder phase. This phase appears when the $\bar{v}$ is very small for all the values of $\varepsilon$ and higher $\bar{v}$ and larger $\varepsilon$. For large $\varepsilon$, particles mostly perform the random action, hence the steady state will be disorder state.   For low values of $\bar{v}$, the spins remain in the same neighbourhood for a long time, and many small clusters are formed, and the distribution of the spin remains nearly homogeneous with a small tail, as shown in Fig. \ref{fig:2}(e).  So, the spins cannot establish any correlation with other spins far away from it, and the system behaves like a collection of non-interacting particles, and we get nearly zero-ordering as shown in Fig.~\ref{fig:2}(a).  When both $\bar{v}$ and $\varepsilon$  are high, spins interact with other spins far from it, but the system remains in disorder because at higher $\varepsilon$, spins take random action with higher probabilities, and interaction is not effective. \\

(ii) {\bf{\textit{Flocking Phase}}}: This phase appears for lower values of $\varepsilon$ and $\bar{v} \sim 0.01$. In the phase diagram, it is shown in a cyan-shaded region. As concluded from Fig.~\ref{fig:2}(f), in this phase, a large cluster is moving in the sea of small clusters spanning over all the space. The motion of these small clusters is quite random in the system, so these clusters perturb the motion of a large cluster. Since the random moving clusters span most of the space, the large cluster is habitual to the perturbation induced by them. Sometimes, these perturbations are able to reverse the direction of motion of big clusters.  \\

(iii) {\bf{\textit{Flipping phase}}}: The green-shaded region in the phase diagram comes when $\bar{v} \sim 0.1$ and  lower $\varepsilon$ range, this indicates the flipping phase. In this phase, the order parameter shows maximum ordering with almost $\pm 1$ value and shows switching from positive to negative and vice versa intermittently, as shown in Fig.~\ref{fig:2}(c). From Fig.~\ref{fig:2}(g), we can conclude that almost all the spins are part of a large cluster, few of them are randomly moving, and most of the space is empty. The small perturbation induced by these few spins can eventually disturb the motion of the cluster and change its direction of motion. The direction change takes place in a few time steps. In Sec.~\ref{III B}, we have discussed this phase in detail and also explained the mechanism of flipping of spins from one sign to another. The flipping of spins is very common in one-dimensional polar flocks, and it has also been reported  in many previous studies \cite{o1999alternating,raymond2006flocking,yates2009inherent,dossetti2011cohesive,benvegnen2022flocking,laighleis2018minimal,pawanDirectionalCue}.\\

(iv) {\bf{\textit{Oscillatory phase}}}: This is a new phase that has not been observed in previous studies of one-dimensional Active Ising spins systems. This phase comes when $(\bar{v} \sim 1)$ and also for the lower $\varepsilon$ range, as shown in the phase diagram using a red-shaded region. In this phase, the whole flock reverses its direction of motion at every time step, which can be seen in the order parameter time series in Fig.~\ref{fig:2}(d), and the distribution of the spins is homogeneous in the system. Detailed discussion about this phase is in Sec.~\ref{III C}. \\

\subsection{Switching Mechanism and Flipping Phase Characterization}\label{III B}

In this section, we study the mechanism of reversing the direction of motion of a single dense cluster in the flipping phase. 
\begin{figure} 
\centering
{\includegraphics[width=0.9 \linewidth]{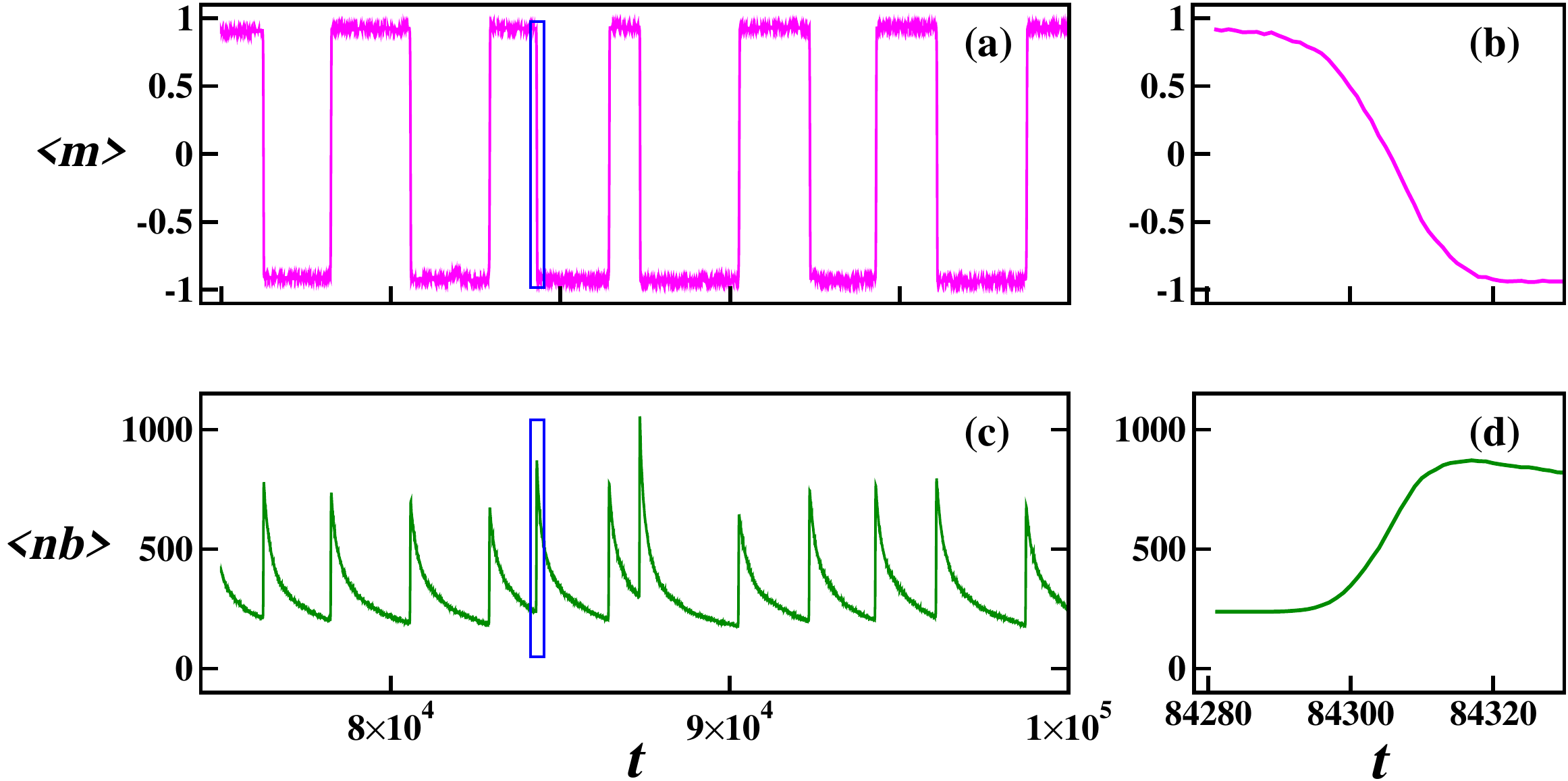}}
    \caption{(color online) Plot (a, c) shows the time series of the order parameter $\langle m \rangle$ and corresponding average number of neighbour spins $\langle nb \rangle $ in the steady state for the Flipping phase. And (c,d) shows the zoomed region of the blue box for $\langle m \rangle$ and $\langle nb \rangle$.  Here $N = 3000, L = 300 \delta x, v_0$ = $0.30$ and $\varepsilon = 0.020$. }
\label{fig:4}
\end{figure}

\begin{figure} 
\centering
{\includegraphics[width=0.85 \linewidth]{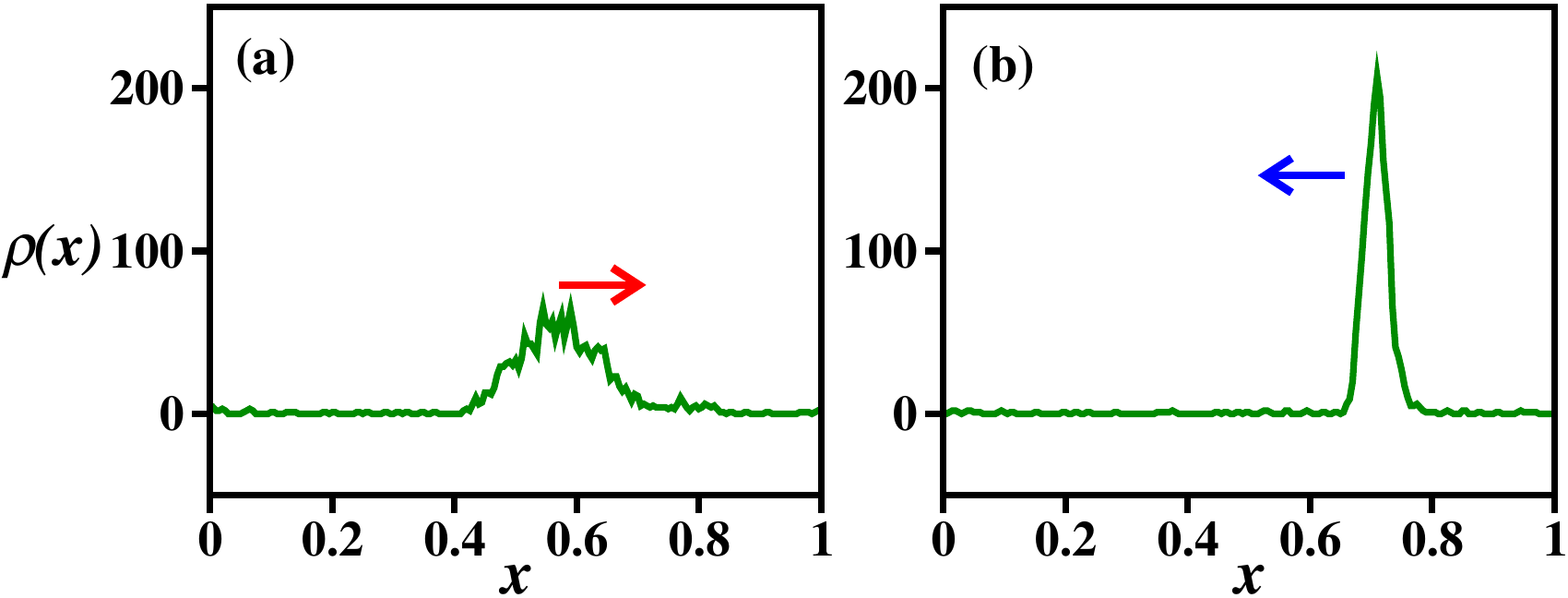}}
    \caption{(color online) The local density $\rho(x)$ profile (a) just before the reversal and (b) just after the reversal of the flock. Just before the reversal, the flock is wider, and after the reversal, it becomes highly dense. The arrow shows the direction of the motion of the flock. Parameters are: $N = 1000$, $L = 100\delta x$, $\varepsilon = 0.020$ and $v_0$ = $0.30$.}
\label{fig:5}
\end{figure}

Fig.~\ref{fig:4}(a) and (c) show the time series of the order parameter $\langle m \rangle$  and the corresponding average number of neighbour spins $\langle nb \rangle $ within the interaction range in the system for the flipping phase in the steady state. The observation of the time series of order parameter suggests that the value of $\langle m \rangle$ intermittently changes from positive to negative, i.e., the flock reverses its direction of motion, as shown in Fig.~\ref{fig:4}(a). The transition from positive to negative value or vice versa happens in a few time steps. In Fig.~\ref{fig:4}(c), the value of $\langle nb \rangle$ is high at the corresponding time at which $\langle m \rangle$ reverses its value from positive to negative or vice versa. So, just after reversal, $\langle nb \rangle$ takes the peak value and slowly decreases until the next reversal happens. In Fig.~\ref{fig:4}(b) and (d), we show the zoom plot of the time series of $\langle m \rangle$ and $\langle nb \rangle$ near one reversal (the blue box shows zoomed region in the time series of $\langle m \rangle$ and $\langle nb \rangle$ ). It is very clear that reversal is almost instantaneous, and as soon as the direction of magnetization reverses, the number of neighbours reaches a peak value simultaneously. The detailed mechanism of reversal is discussed below. \\

In Fig.~\ref{fig:5}(a-b), the local density profile $\rho(x)$ of the system is shown (a) just before the reversal, (b) after the reversal. The local density $\rho(x)$ is calculated by dividing the whole system into bins and calculating the number of spins in these regions.
In Fig.~\ref{fig:5}(a), the local density $\rho(x)$ profile of the flock (moving right) just before the starting of the reversal is shown. Here, the maximum value of $\rho(x)$ is less, and the flock has a wider shape. This wider flock undergo a stochastic reversal and again becomes dense. In Fig.~\ref{fig:5}(b), just after the reversal, the value of $\rho$ is high, i.e. most of the spins are condensed in a small region, so there is a single dense cluster (moving left).

\begin{figure*} 
\centering
{\includegraphics[width=0.8\linewidth]{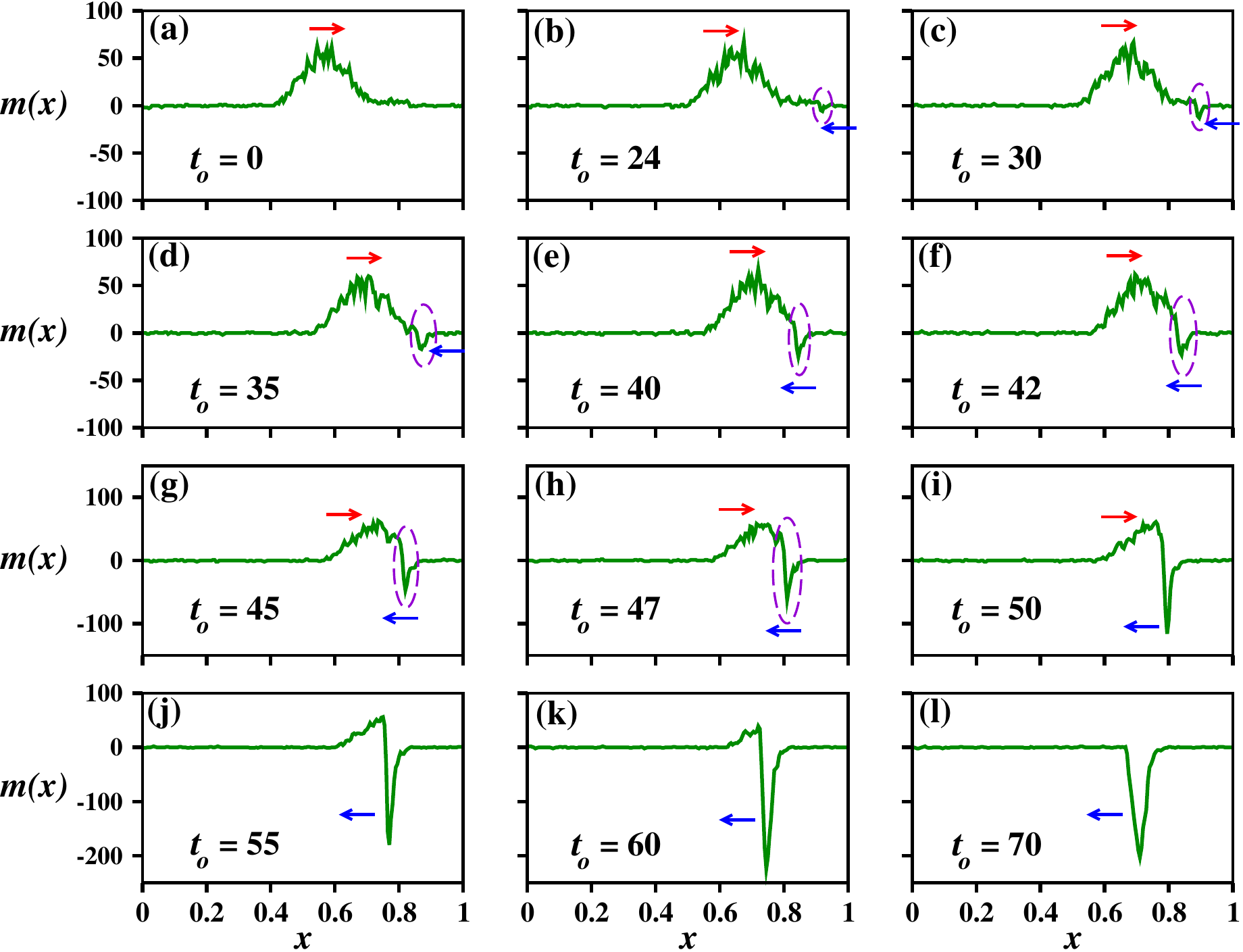}}
    \caption{(color online) Plot (a-l) shows the local magnetization $m(x)$ profile of the system during the reversal of the direction of the flock from right to left. The red and blue arrows represent the flock moving to the right and left, respectively. Initially, the whole flock (wider) moves towards the right (most of the spins have $+1$ orientation) (a) a cluster of very few spins moving to the left appears (b) this small cluster collides with the front of the large cluster (c) It starts to change the orientation of its neighbour spins (d) it grows with the sharp peak (e-f) All spins gradually have their orientation flipped as it spreads across the flock. The peak value of $m(x)$ for it is large, and the width is much smaller than the large flock (g-k). Finally, the orientation of all the spins changes, and a condensed flock moves to the left (most of the spins have $-1$ orientation) (l). All the parameters are similar to the Fig.~\ref{fig:5}.}
\label{fig:6}
\end{figure*}

\begin{figure*} 
\centering
{\includegraphics[width=0.8 \linewidth]{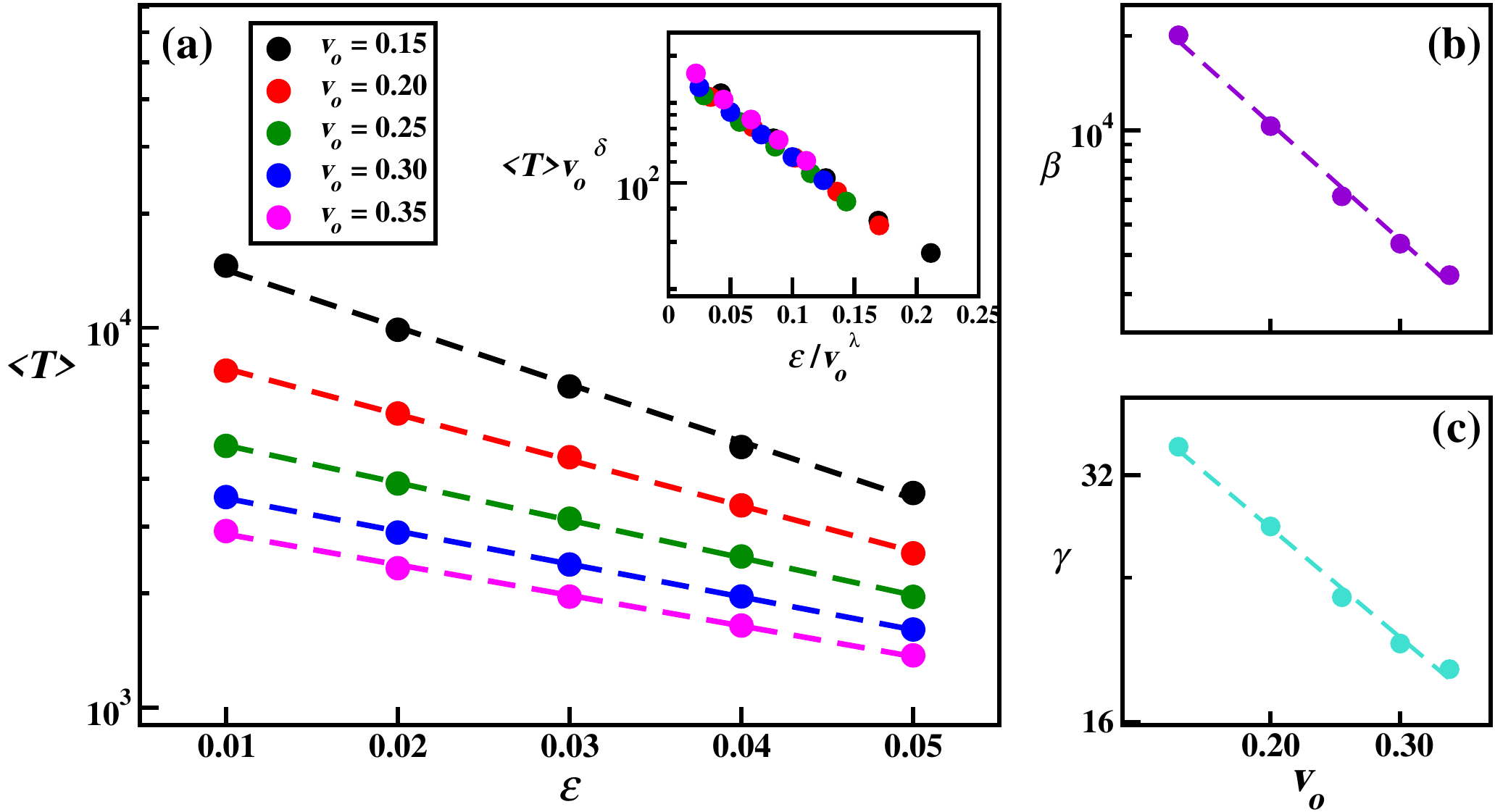}}
    \caption{(color online) Plot (a) shows the variation in mean reversal time $\langle T \rangle$  with respect to exploration probability $\varepsilon$ for different mean self-propulsion speeds $v_0$ on a semi-log y scale. \textit{Inset}: The plot shows Scaled $\langle T \rangle$ vs. Scaled $\varepsilon$ for different $v_0$.  Plots (b) and (c) show the variation of $\beta$ and $\gamma$ with $v_0$ on a log-log scale. The symbols show the data points, and the dashed lines show the fitting. All other parameters are similar to Fig.~\ref{fig:2}.}
\label{fig:7}
\end{figure*}

\begin{figure} 
\centering
{\includegraphics[width=0.95 \linewidth]{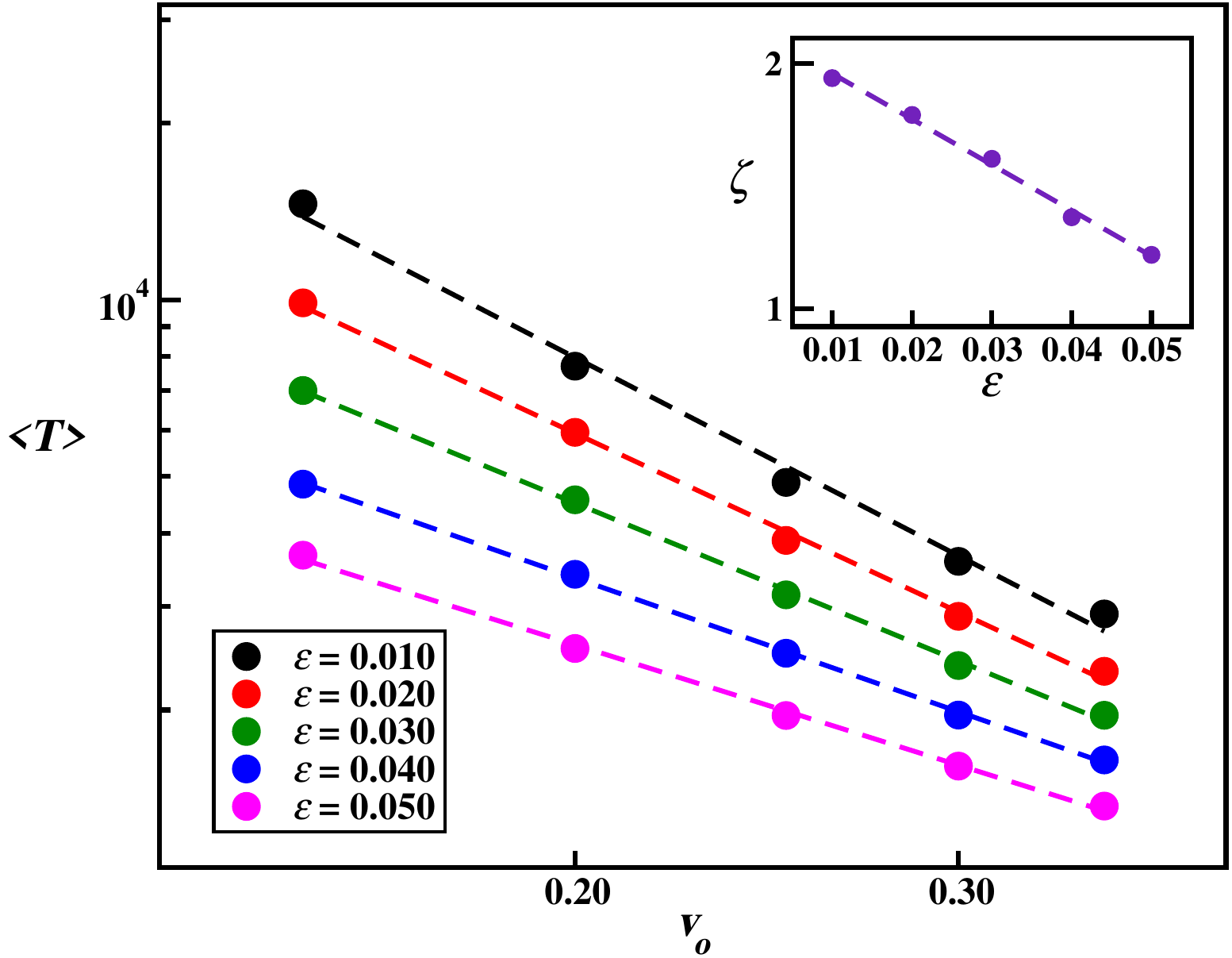}}
    \caption{(color online) Plot shows the variation of mean reversal time $\langle T \rangle$  with respect to mean self-propulsion speed $v_0$ for different exploration probabilities $\varepsilon$ on a log-log scale. \textit{Inset}: the variation of the power $\zeta$ with $\varepsilon$ is shown on a semi-log y scale. The symbols show the data points, and the dashed lines show the fitting. All other parameters are similar to Fig. \ref{fig:7}.}
\label{fig:8}
\end{figure}

In Fig.~\ref{fig:6}(a-l), the time evolution of the local magnetization $m(x)$ during the reversal of the flock's direction from right to left is shown. Just after the reversal, the cluster is dense and then slowly dilutes, as shown in Fig.~\ref{fig:5}(b) and (a), respectively. The time $t_o$ has been counted in steady-state when we started the observation.  In Fig.~\ref{fig:6}(a), at $t_o = 0$, a single wide cluster is moving towards the right (almost all the spins have $+1$ orientation). 
In the studies of one-dimension systems \cite{solon2013revisiting}, it has been found that when a cluster moves in the system, the velocity of the fore-end is higher than the back end until the cluster spreads uniformly. So, from the fore-end, it has an extended tail, then the back-end. A tiny cluster with a small number of spins moving towards the left appears at the fore-end of the large cluster at $t_o = 24$ [Fig.~\ref{fig:6}(b)]. The large cluster collides with the small cluster from the fore-end [Fig.~\ref{fig:6}(c)]. The peak value of $m(x)$ for the small cluster is larger than the large cluster at the fore-end. So, near this peak, the spins that are moving right start to change their orientation from $+1$ to $-1$. Since the fore-end has higher velocity, the peak amplitude of the small left-moving cluster increases rapidly. The width of the left moving cluster is narrow, and it is maintained during the reversal process. So, the left moving cluster takes over the right moving cluster in a few time steps [Fig.~\ref{fig:6}(d-k)]. Finally, there is a single, dense, narrow, left-moving cluster where almost all the spins have $-1$ orientation [Fig.~\ref{fig:6}(l)]. The amplitude of the peak value of $m(x)$ is very high for this left-moving cluster in Fig.~\ref{fig:6}(l) compared to the right-moving cluster in Fig.~\ref{fig:6}(a). So, a small left-moving cluster appears in the fore-end of a right-moving large cluster, and the small cluster starts to change the orientation of spins of the large cluster from $+1$ to $-1$. In a few time steps, this small cluster takes over the large cluster, and a dense cluster appears, moving towards the left. After a long time, this cluster also dilutes, and due to a right-moving small cluster, it turns into a right-moving cluster. This process keeps repeating in steady state. In this phase, for the formation of a single flock or to maintain cohesion, the reversal of the direction of the flock is necessary \cite{o1999alternating,sakaguchi2019flip}.\\

To check how frequently the flock reverses its direction of motion, we also calculate the mean reversal time $\langle T \rangle$ (The average time for the flock to reverse its direction of motion in the steady state). In Fig.~\ref{fig:7}(a), the mean reversal time $\langle T \rangle$  vs. exploration probability $\varepsilon$ for different mean self-propulsion speeds $v_0$, is shown on a semi-$\log y$ scale. As $\varepsilon$ increases, $\langle T \rangle$ decays for fixed $v_0$. Increasing $v_0$ leads to slower decay of $\langle T \rangle$. As the value of $\varepsilon$ increases, the probability of taking the random action increases, which can be considered as an increment in the noise that leads to more fluctuations in the system. Due to these fluctuations, the frequency of reversing the direction of motion of the flock increases, i.e. $\langle T \rangle$ decreases as $\varepsilon$ increases.  In Fig.~\ref{fig:7}(a), data points have been obtained from simulation, and dashed lines show exponential fit. So $\langle T \rangle = \beta e^{-\gamma \varepsilon}$. From exponential fitting, we have extracted $\beta$ and $\gamma$ for different $v_0$. In Fig.~\ref{fig:7}(b-c), we have shown the variation of $\beta$ and $\gamma$ with $v_0$ on the $\log-\log$ scale. The data for mean reversal time $\langle T \rangle$ shows data collapse when we rescale the $\langle T \rangle$ on the $y-$ axis with a power of $v_0^{\delta}$ and $\varepsilon$ parameter on the $x-$ axis with another power of $v_0^{-\lambda}$. Where the $\delta$ and $\lambda$ are obtained from the power law fitting of $\beta$ and $\gamma$ in the Fig.~\ref{fig:7}(b-c), respectively. We have found that $\delta = 2.10 \pm 0.09$ and $\lambda = 0.76 \pm 0.03$. The data collapse suggests the invariance of mean reversal time $\langle T \rangle$ over different mean self-propulsion speeds $v_0$.  \\

In Fig.~\ref{fig:8}, the variation of mean reversal time $\langle T \rangle$  with respect to the  $v_0$ for different exploration probabilities $\varepsilon$ is shown on a $\log-\log$ scale. For a fixed $\varepsilon$, $\langle T \rangle$ decays as a power law as $v_0$ increases. The solid circles are data points, and the dashed lines show a power-law fit. The large $v_0$ leads to the more frequent flipping of the flock: since larger $v_0$ means bigger hopping steps of particles. Hence it increases the probability of interaction of particles from two different clusters. As we have explained in the previous paragraph and shown in Fig.~\ref{fig:6}(a-l), the larger steps can result in a more frequent collision of two oppositely moving clusters that increase the flipping probability, and hence the flipping time decreases. $\langle T \rangle$ shows the power law decay with increasing $v_0$.  $\langle T \rangle = B v_0^{\zeta}$. In the inset, the decrease of the slope $\zeta$ with respect to $\varepsilon$ is shown on a semi-$\log y$ scale. The dashed line shows exponential fit for all data points. Unlike the dependence of exponents in the previous plot~\ref{fig:7}(b-c) with respect to $v_0$ is a power law, here we find the $\zeta$ dependence on $\varepsilon$ as exponential.\\
\subsection{Oscillatory Phase: A Chaotic phase}\label{III C}
\begin{figure} 
\centering
{\includegraphics[width=0.95 \linewidth]{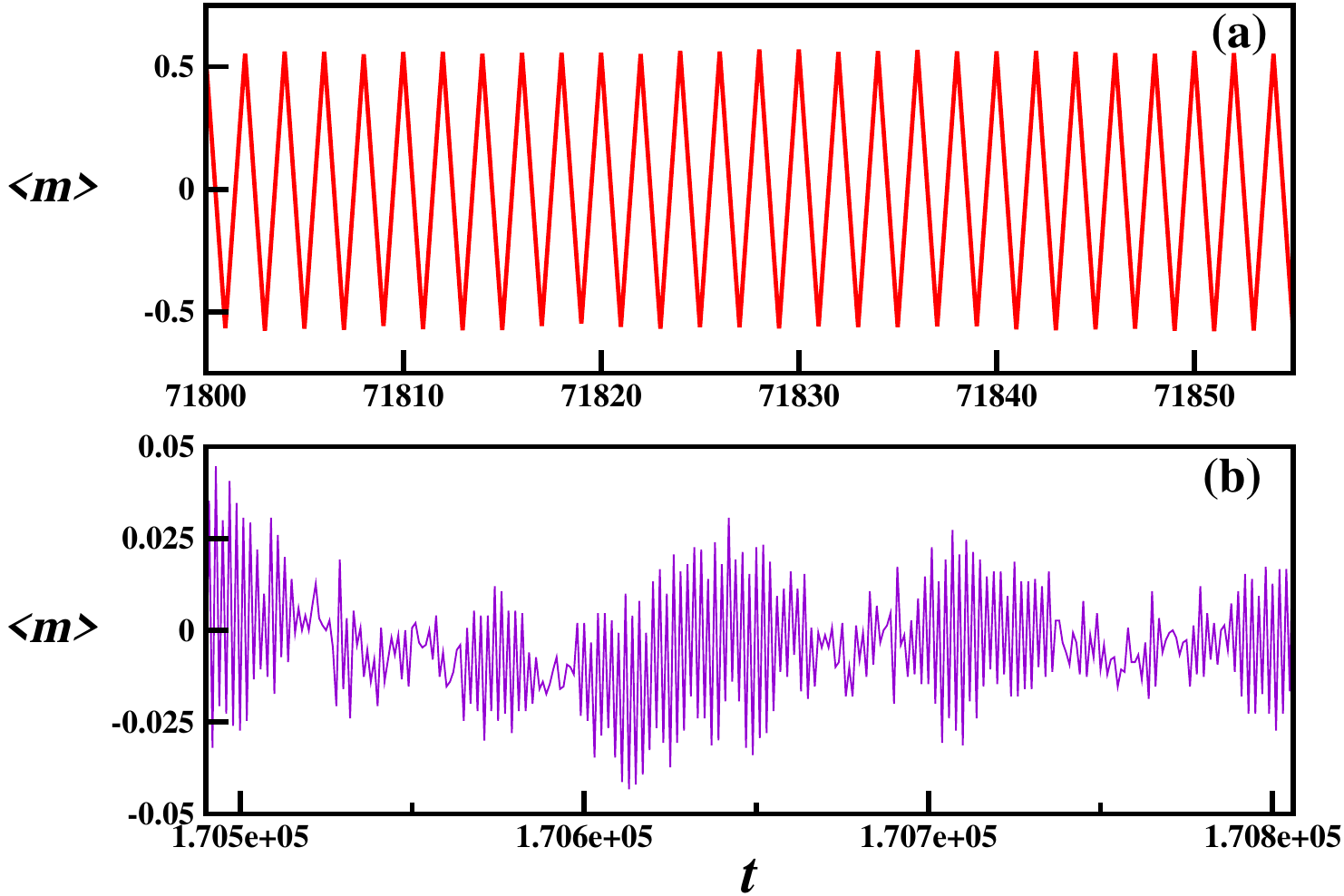}}
    \caption{(color online) Plot (a) shows the periodic oscillation part of the order parameter time series. Plot (b) indicates the Spatiotemporal intermittency (SPI) part in the time series where the order parameter is very small. All other parameters are similar to Fig.~\ref{fig:2}(d).}
\label{fig:9}
\end{figure}

Here, we discuss the new oscillatory phase in detail. This phase appears when the self-propulsion speed $\bar{v}$ and $\delta x/\delta t$ are comparable and exploration probability $\varepsilon < 0.30$. In this phase, the time series of order parameter $\langle m \rangle$ shows periodic oscillation with a finite ordering followed by Spatiotemporal intermittency (STI) where the ordering decreases, sometimes approaching zero as shown in Fig.~\ref{fig:9}(a) and (b), respectively. This STI persists for a finite time period before the system returns to periodic oscillations. This cycle of ordered oscillation transitioning to STI and back to ordered oscillation recurs repeatedly throughout the time series in the steady state. In the oscillation state, the system shows a long-range correlation, and in STI, it decays rapidly. The distribution of the spins in the system is homogeneous, as shown in Fig. \ref{fig:2}(h). Since the order of $\delta x/\delta t$ and $\bar{v}$ is equivalent, in a single time step, spins can go to the new neighbourhood, and they get a small chance to interact with the neighbouring spins and since they keep hopping from one cluster to another and their orientation keep switching. This is true for all the spins, and finally, it results in a small global order parameter and oscillation of the mean order parameter at every time step. \\

\begin{figure} 
\centering
{\includegraphics[width=0.85 \linewidth]{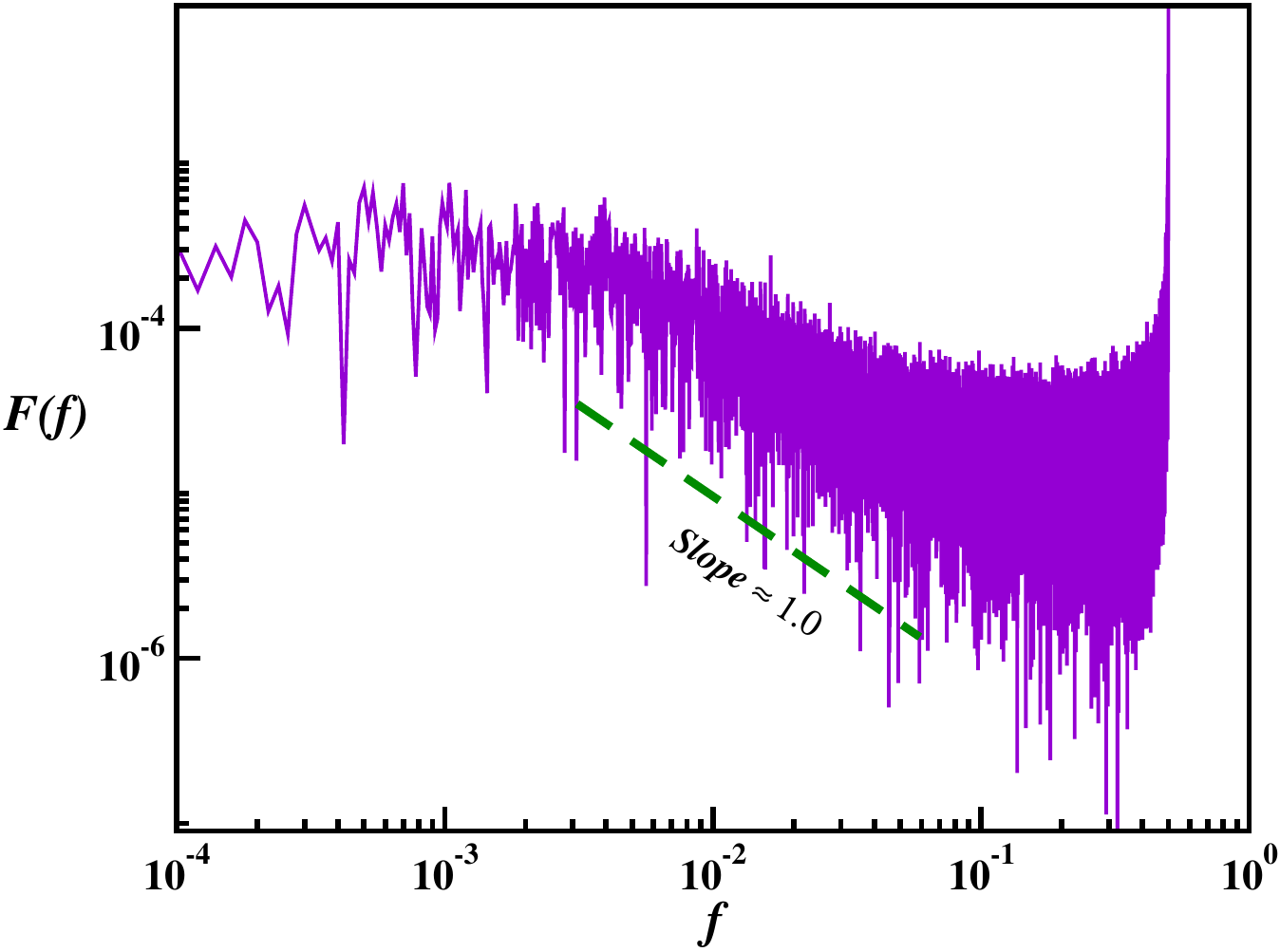}}
    \caption{(color online) Plot shows the FFT spectrum of order parameter time series in the steady state. The spectrum is continuous with an intense peak at $f=0.5$. All parameters are similar to Fig.~\ref{fig:2}(d).}
\label{fig:10}
\end{figure}

\begin{figure*}
\centering
{\includegraphics[width=0.85 \linewidth]{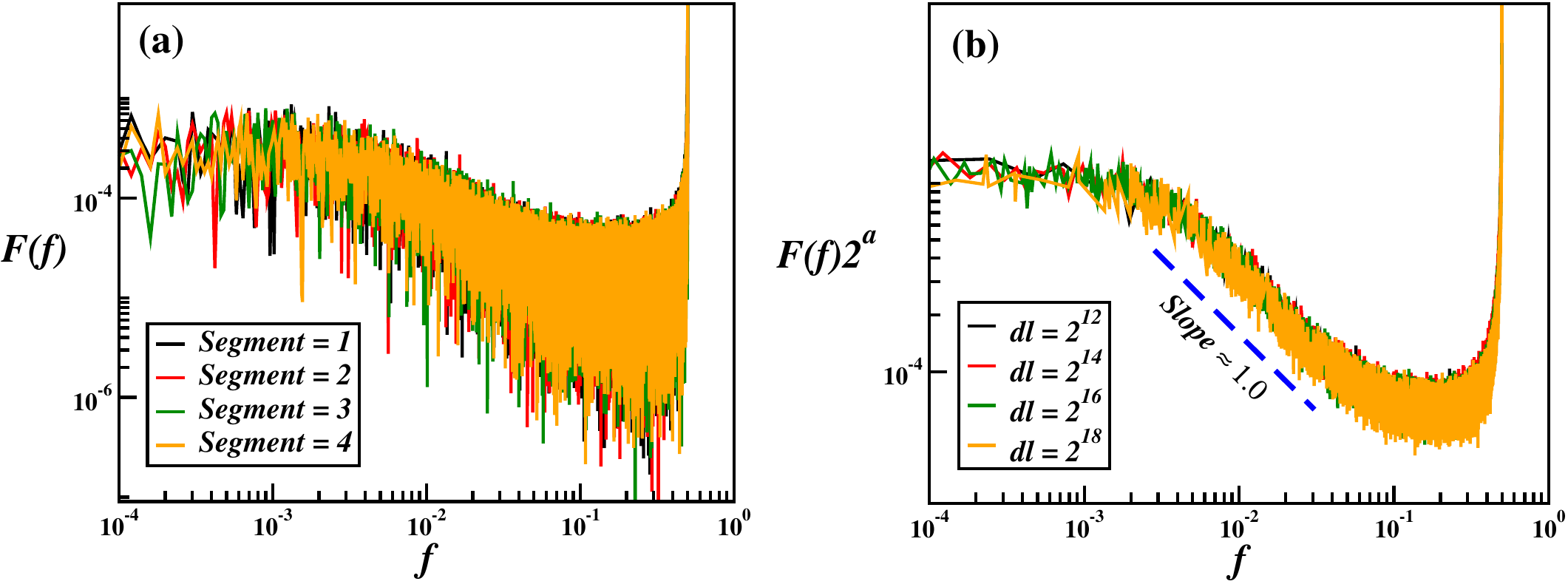}}
    \caption{(color online) Plot (a) shows the stationarity of the Fourier spectrum for four different segments of the order parameter time series. Plot (b) illustrates the data collapse of the Fourier spectrum for four time series, each of increasing length: $2^n$ where $n = 12,14,16,18$. The vertical axis is scaled using the parameter $a = 0.5(n-n_o)$, with $n_o = 12$. All parameters are similar to Fig.~\ref{fig:2}(d).}
\label{fig:11}
\end{figure*}

To characterize this phase, we have calculated the Fast Fourier Transform (FFT) spectrum of the time series of order parameter. We have found that its FFT spectrum $F(f)$ shows a continuous spectrum with an intense peak at frequency $f=0.5$, as depicted in Fig.~\ref{fig:10}. This peak indicates the existence of periodic oscillation with a period of 2-time steps in the time series. At intermediate frequencies, it shows power law decay. The contribution comes from STI, showing a continuous range of frequencies.\\
Here we have used the Fourier transform technique for spectral analysis, which requires the stationarity of the Fourier spectrum over time. To check the same, in the steady state of the order parameter time series, it has been divided into four distinct segments, each consisting of $50,000$-time steps, and calculate the FFT spectrum for each segment. We observe that for all the segments, all the curves collapse nicely, as shown in Fig.~\ref{fig:11}(a). It proves the stationarity. \\
To check the dependence of $F(f)$ on the finite size of the time series, the time series data is segmented into four distinct parts, each starting from time $t'$ in steady state. The length of each segment $dl$ increases in powers of 2, with the first segment having a length of $2^{12}$ time steps. The subsequent segments are progressively longer, with lengths of $2^{14}, 2^{16}$, and $2^{18}$ respectively. For each segment of different length, we have calculated the FFT spectrum $F(f)$ and averaged it over $30$ different realizations. We have found that as the length of segment $dl$ increases, the $F(f)$ shifts towards the lower values. And, it scales as $1/{dl}$ at intermediate frequencies. \cite{valsakumar1997signature}. That's why in Fig.~\ref{fig:11}(b), the data shows good collapse. Since the computed spectrum differs from the actual spectrum because of the finite size of the time series, it vanishes as $dl \rightarrow \infty$.
As suggested in \cite{maryshev2019dry,strogatz2018nonlinear,valsakumar1997signature}, continuous spectrum and its dependence on the data length are signatures of chaotic dynamics.\\ 
One more quantity that supports chaotic dynamics is the positive Lyapunov exponent \cite{wolf1985determining,strogatz2018nonlinear}. Using the Wolf algorithm \cite{wolf1985determining, kodba2004detecting}, we have calculated the largest Lyapunov exponent $\Lambda_{max}$ for the time series of the order parameter. The largest Lyapunov exponent $\Lambda_{max}$ for a time series is defined as:
\begin{equation}
    \Lambda_{max} = \frac{1}{Mt_{evolv}} \sum_{i=0}^{M} ln\frac{L^{(i)}_{evolv}}{L^{(i)}_0}
\end{equation}
First, using time-delay embedding, the phase space is reconstructed for the time series and finds the nearest neighbour. For two nearest neighbour points in reconstructed phase space, $L_0$ is an initial distance, and $L_{evolv}$ is the final distance after a fixed evolution time $t_{evolv}$. $M$ is the total number of replacement steps.
All these details regarding the calculation of $\Lambda_{max}$ using numerical simulation are out of the scope of this paper. We have used the approach for the calculation of the Lyapunov exponent for a time series, which is nicely explained in \cite{kodba2004detecting}.\\
In Table-~\ref{Table:1}, we have shown the largest Lyapunov exponent $\Lambda_{max}$ for the order parameter time series of different self-propulsion speeds. The positive Lyapunov exponents suggest that the oscillatory phase is chaotic in nature.

\begin{table}[h!]
\centering
\begin{tabular}{|c|c|}
\hline
$v_0$ & $\Lambda_{max} \pm  \Delta \Lambda_{max}$ \\ \hline
2   &   $0.0385   \pm  0.0049$    \\ \hline
3   &   $0.0376   \pm  0.0082 $    \\ \hline
4   &   $0.0368   \pm  0.0095$   \\ \hline
\end{tabular}
\caption{The Table shows variation in the largest Lyapunov exponent with respect to mean self-propulsion speed in the oscillatory phase.}
\label{Table:1}
\end{table}

\section{Conclusions}\label{IV}
Using the Reinforcement Learning approach, we have developed an algorithm to study the flocking in a one-dimensional system with active Ising spins. We have explored the system for a wide range of self-propulsion speeds and exploration parameter $(\varepsilon)$. At a very low self-propulsion speed, the system is in the disorder phase with no ordering. As speed continuously increases, the system goes to flocking and then flipping, where the system reverses its direction of motion intermittently. These are the phases that have already been reported in many previous studies. We utilized the $RL$ approach and explained the key characteristics of different phases. Except for these phases, we have found a new phase where the system reverses its direction of motion at each instant; we named it the oscillatory phase. It appears when the hopping steps are of the order of the interaction range. In this work, we have thoroughly investigated the mechanism and characteristics of the flipping phase. In this phase, the system reverses its direction of motion due to a small cluster of opposite-moving spins. The mean reversal time decays exponentially with exploration probability and shows the scaling. For the oscillatory phase, the Fourier spectrum and positive Lyapunov exponent suggest that it is chaotic. \\
The outcomes of the $RL$ technique also closely mirror those of other traditional one-dimensional flocking investigations. We have created an $RL$ framework in this work that can be applicable to flocking systems of larger dimensions as well. In the present study, the learning is utilized only in the orientation updates of the spins, where spins learn by maintaining cohesion among them. The learning outcome, along with other means, like the step size of the particle hopping, can be interesting to explore in future works.\\

\section{Data availability statement}
The data are available upon reasonable request from the authors.\\
\section{acknowledgment}\label{V}
A.K., P.K.M., P.S.M., R.S., S.M. and D.G. thank PARAM Shivay for the computational facility under the National Supercomputing Mission, Government of India, at the Indian Institute of Technology (BHU) Varanasi and also I.I.T. (BHU) Varanasi computational facility. A. K. thanks PMRF, INDIA, for the research fellowship. S. M. thanks DST, SERB (INDIA), Project No. CRG/2021/006945 and MTR/2021/000438 for partial financial support.\\
\section{Conflict of Interest declaration}
The authors declare that there are no conflicts of interest.


\begin{thebibliography}{63}%
\makeatletter
\providecommand \@ifxundefined [1]{%
 \@ifx{#1\undefined}
}%
\providecommand \@ifnum [1]{%
 \ifnum #1\expandafter \@firstoftwo
 \else \expandafter \@secondoftwo
 \fi
}%
\providecommand \@ifx [1]{%
 \ifx #1\expandafter \@firstoftwo
 \else \expandafter \@secondoftwo
 \fi
}%
\providecommand \natexlab [1]{#1}%
\providecommand \enquote  [1]{``#1''}%
\providecommand \bibnamefont  [1]{#1}%
\providecommand \bibfnamefont [1]{#1}%
\providecommand \citenamefont [1]{#1}%
\providecommand \href@noop [0]{\@secondoftwo}%
\providecommand \href [0]{\begingroup \@sanitize@url \@href}%
\providecommand \@href[1]{\@@startlink{#1}\@@href}%
\providecommand \@@href[1]{\endgroup#1\@@endlink}%
\providecommand \@sanitize@url [0]{\catcode `\\12\catcode `\$12\catcode `\&12\catcode `\#12\catcode `\^12\catcode `\_12\catcode `\%12\relax}%
\providecommand \@@startlink[1]{}%
\providecommand \@@endlink[0]{}%
\providecommand \url  [0]{\begingroup\@sanitize@url \@url }%
\providecommand \@url [1]{\endgroup\@href {#1}{\urlprefix }}%
\providecommand \urlprefix  [0]{URL }%
\providecommand \Eprint [0]{\href }%
\providecommand \doibase [0]{http://dx.doi.org/}%
\providecommand \selectlanguage [0]{\@gobble}%
\providecommand \bibinfo  [0]{\@secondoftwo}%
\providecommand \bibfield  [0]{\@secondoftwo}%
\providecommand \translation [1]{[#1]}%
\providecommand \BibitemOpen [0]{}%
\providecommand \bibitemStop [0]{}%
\providecommand \bibitemNoStop [0]{.\EOS\space}%
\providecommand \EOS [0]{\spacefactor3000\relax}%
\providecommand \BibitemShut  [1]{\csname bibitem#1\endcsname}%
\let\auto@bib@innerbib\@empty
\bibitem [{\citenamefont {Cichos}\ \emph {et~al.}(2020)\citenamefont {Cichos}, \citenamefont {Gustavsson}, \citenamefont {Mehlig},\ and\ \citenamefont {Volpe}}]{cichos2020machine}%
  \BibitemOpen
  \bibfield  {author} {\bibinfo {author} {\bibfnamefont {F.}~\bibnamefont {Cichos}}, \bibinfo {author} {\bibfnamefont {K.}~\bibnamefont {Gustavsson}}, \bibinfo {author} {\bibfnamefont {B.}~\bibnamefont {Mehlig}}, \ and\ \bibinfo {author} {\bibfnamefont {G.}~\bibnamefont {Volpe}},\ }\href@noop {} {\bibfield  {journal} {\bibinfo  {journal} {Nature Machine Intelligence}\ }\textbf {\bibinfo {volume} {2}},\ \bibinfo {pages} {94} (\bibinfo {year} {2020})}\BibitemShut {NoStop}%
\bibitem [{\citenamefont {dos Santos~Mignon}\ and\ \citenamefont {da~Rocha}(2017)}]{dos2017adaptive}%
  \BibitemOpen
  \bibfield  {author} {\bibinfo {author} {\bibfnamefont {A.}~\bibnamefont {dos Santos~Mignon}}\ and\ \bibinfo {author} {\bibfnamefont {R.~L. d.~A.}\ \bibnamefont {da~Rocha}},\ }\href@noop {} {\bibfield  {journal} {\bibinfo  {journal} {Procedia Computer Science}\ }\textbf {\bibinfo {volume} {109}},\ \bibinfo {pages} {1146} (\bibinfo {year} {2017})}\BibitemShut {NoStop}%
\bibitem [{\citenamefont {Canese}\ \emph {et~al.}(2021)\citenamefont {Canese}, \citenamefont {Cardarilli}, \citenamefont {Di~Nunzio}, \citenamefont {Fazzolari}, \citenamefont {Giardino}, \citenamefont {Re},\ and\ \citenamefont {Span{\`o}}}]{canese2021multi}%
  \BibitemOpen
  \bibfield  {author} {\bibinfo {author} {\bibfnamefont {L.}~\bibnamefont {Canese}}, \bibinfo {author} {\bibfnamefont {G.~C.}\ \bibnamefont {Cardarilli}}, \bibinfo {author} {\bibfnamefont {L.}~\bibnamefont {Di~Nunzio}}, \bibinfo {author} {\bibfnamefont {R.}~\bibnamefont {Fazzolari}}, \bibinfo {author} {\bibfnamefont {D.}~\bibnamefont {Giardino}}, \bibinfo {author} {\bibfnamefont {M.}~\bibnamefont {Re}}, \ and\ \bibinfo {author} {\bibfnamefont {S.}~\bibnamefont {Span{\`o}}},\ }\href@noop {} {\bibfield  {journal} {\bibinfo  {journal} {Applied Sciences}\ }\textbf {\bibinfo {volume} {11}},\ \bibinfo {pages} {4948} (\bibinfo {year} {2021})}\BibitemShut {NoStop}%
\bibitem [{\citenamefont {Hook}\ \emph {et~al.}(2021)\citenamefont {Hook}, \citenamefont {El-Sedky}, \citenamefont {De~Silva},\ and\ \citenamefont {Kondoz}}]{hook2021learning}%
  \BibitemOpen
  \bibfield  {author} {\bibinfo {author} {\bibfnamefont {J.}~\bibnamefont {Hook}}, \bibinfo {author} {\bibfnamefont {S.}~\bibnamefont {El-Sedky}}, \bibinfo {author} {\bibfnamefont {V.}~\bibnamefont {De~Silva}}, \ and\ \bibinfo {author} {\bibfnamefont {A.}~\bibnamefont {Kondoz}},\ }\href@noop {} {\bibfield  {journal} {\bibinfo  {journal} {Cognitive Systems Research}\ }\textbf {\bibinfo {volume} {65}},\ \bibinfo {pages} {40} (\bibinfo {year} {2021})}\BibitemShut {NoStop}%
\bibitem [{\citenamefont {Sutton}\ and\ \citenamefont {Barto}(2018)}]{sutton2018reinforcement}%
  \BibitemOpen
  \bibfield  {author} {\bibinfo {author} {\bibfnamefont {R.~S.}\ \bibnamefont {Sutton}}\ and\ \bibinfo {author} {\bibfnamefont {A.~G.}\ \bibnamefont {Barto}},\ }\href@noop {} {\emph {\bibinfo {title} {Reinforcement learning: An introduction}}}\ (\bibinfo  {publisher} {MIT press},\ \bibinfo {year} {2018})\BibitemShut {NoStop}%
\bibitem [{\citenamefont {Watkins}\ and\ \citenamefont {Dayan}(1992)}]{watkins1992q}%
  \BibitemOpen
  \bibfield  {author} {\bibinfo {author} {\bibfnamefont {C.~J.}\ \bibnamefont {Watkins}}\ and\ \bibinfo {author} {\bibfnamefont {P.}~\bibnamefont {Dayan}},\ }\href@noop {} {\bibfield  {journal} {\bibinfo  {journal} {Machine learning}\ }\textbf {\bibinfo {volume} {8}},\ \bibinfo {pages} {279} (\bibinfo {year} {1992})}\BibitemShut {NoStop}%
\bibitem [{\citenamefont {Mohan}\ \emph {et~al.}(2021)\citenamefont {Mohan}, \citenamefont {Sharma},\ and\ \citenamefont {Narayan}}]{mohan2021optimal}%
  \BibitemOpen
  \bibfield  {author} {\bibinfo {author} {\bibfnamefont {P.}~\bibnamefont {Mohan}}, \bibinfo {author} {\bibfnamefont {L.}~\bibnamefont {Sharma}}, \ and\ \bibinfo {author} {\bibfnamefont {P.}~\bibnamefont {Narayan}},\ }in\ \href@noop {} {\emph {\bibinfo {booktitle} {2021 5th International Conference on Intelligent Computing and Control Systems (ICICCS)}}}\ (\bibinfo {organization} {IEEE},\ \bibinfo {year} {2021})\ pp.\ \bibinfo {pages} {811--817}\BibitemShut {NoStop}%
\bibitem [{\citenamefont {Schulman}\ \emph {et~al.}(2017)\citenamefont {Schulman}, \citenamefont {Wolski}, \citenamefont {Dhariwal}, \citenamefont {Radford},\ and\ \citenamefont {Klimov}}]{schulman2017proximal}%
  \BibitemOpen
  \bibfield  {author} {\bibinfo {author} {\bibfnamefont {J.}~\bibnamefont {Schulman}}, \bibinfo {author} {\bibfnamefont {F.}~\bibnamefont {Wolski}}, \bibinfo {author} {\bibfnamefont {P.}~\bibnamefont {Dhariwal}}, \bibinfo {author} {\bibfnamefont {A.}~\bibnamefont {Radford}}, \ and\ \bibinfo {author} {\bibfnamefont {O.}~\bibnamefont {Klimov}},\ }\href@noop {} {\bibfield  {journal} {\bibinfo  {journal} {arXiv preprint arXiv:1707.06347}\ } (\bibinfo {year} {2017})}\BibitemShut {NoStop}%
\bibitem [{\citenamefont {Pin{\c{c}}e}\ \emph {et~al.}(2016)\citenamefont {Pin{\c{c}}e}, \citenamefont {Velu}, \citenamefont {Callegari}, \citenamefont {Elahi}, \citenamefont {Gigan}, \citenamefont {Volpe},\ and\ \citenamefont {Volpe}}]{pincce2016disorder}%
  \BibitemOpen
  \bibfield  {author} {\bibinfo {author} {\bibfnamefont {E.}~\bibnamefont {Pin{\c{c}}e}}, \bibinfo {author} {\bibfnamefont {S.~K.}\ \bibnamefont {Velu}}, \bibinfo {author} {\bibfnamefont {A.}~\bibnamefont {Callegari}}, \bibinfo {author} {\bibfnamefont {P.}~\bibnamefont {Elahi}}, \bibinfo {author} {\bibfnamefont {S.}~\bibnamefont {Gigan}}, \bibinfo {author} {\bibfnamefont {G.}~\bibnamefont {Volpe}}, \ and\ \bibinfo {author} {\bibfnamefont {G.}~\bibnamefont {Volpe}},\ }\href@noop {} {\bibfield  {journal} {\bibinfo  {journal} {Nature communications}\ }\textbf {\bibinfo {volume} {7}},\ \bibinfo {pages} {10907} (\bibinfo {year} {2016})}\BibitemShut {NoStop}%
\bibitem [{\citenamefont {Walther}\ and\ \citenamefont {M{\"u}ller}(2008)}]{walther2008janus}%
  \BibitemOpen
  \bibfield  {author} {\bibinfo {author} {\bibfnamefont {A.}~\bibnamefont {Walther}}\ and\ \bibinfo {author} {\bibfnamefont {A.~H.}\ \bibnamefont {M{\"u}ller}},\ }\href@noop {} {\bibfield  {journal} {\bibinfo  {journal} {Soft matter}\ }\textbf {\bibinfo {volume} {4}},\ \bibinfo {pages} {663} (\bibinfo {year} {2008})}\BibitemShut {NoStop}%
\bibitem [{\citenamefont {Hu}\ \emph {et~al.}(2012)\citenamefont {Hu}, \citenamefont {Zhou}, \citenamefont {Sun}, \citenamefont {Fang},\ and\ \citenamefont {Wu}}]{hu2012fabrication}%
  \BibitemOpen
  \bibfield  {author} {\bibinfo {author} {\bibfnamefont {J.}~\bibnamefont {Hu}}, \bibinfo {author} {\bibfnamefont {S.}~\bibnamefont {Zhou}}, \bibinfo {author} {\bibfnamefont {Y.}~\bibnamefont {Sun}}, \bibinfo {author} {\bibfnamefont {X.}~\bibnamefont {Fang}}, \ and\ \bibinfo {author} {\bibfnamefont {L.}~\bibnamefont {Wu}},\ }\href@noop {} {\bibfield  {journal} {\bibinfo  {journal} {Chemical Society Reviews}\ }\textbf {\bibinfo {volume} {41}},\ \bibinfo {pages} {4356} (\bibinfo {year} {2012})}\BibitemShut {NoStop}%
\bibitem [{\citenamefont {Zhang}\ \emph {et~al.}(2017)\citenamefont {Zhang}, \citenamefont {Grzybowski},\ and\ \citenamefont {Granick}}]{zhang2017janus}%
  \BibitemOpen
  \bibfield  {author} {\bibinfo {author} {\bibfnamefont {J.}~\bibnamefont {Zhang}}, \bibinfo {author} {\bibfnamefont {B.~A.}\ \bibnamefont {Grzybowski}}, \ and\ \bibinfo {author} {\bibfnamefont {S.}~\bibnamefont {Granick}},\ }\href@noop {} {\bibfield  {journal} {\bibinfo  {journal} {Langmuir}\ }\textbf {\bibinfo {volume} {33}},\ \bibinfo {pages} {6964} (\bibinfo {year} {2017})}\BibitemShut {NoStop}%
\bibitem [{\citenamefont {Wang}\ \emph {et~al.}(2015)\citenamefont {Wang}, \citenamefont {Duan}, \citenamefont {Ahmed}, \citenamefont {Sen},\ and\ \citenamefont {Mallouk}}]{wang2015one}%
  \BibitemOpen
  \bibfield  {author} {\bibinfo {author} {\bibfnamefont {W.}~\bibnamefont {Wang}}, \bibinfo {author} {\bibfnamefont {W.}~\bibnamefont {Duan}}, \bibinfo {author} {\bibfnamefont {S.}~\bibnamefont {Ahmed}}, \bibinfo {author} {\bibfnamefont {A.}~\bibnamefont {Sen}}, \ and\ \bibinfo {author} {\bibfnamefont {T.~E.}\ \bibnamefont {Mallouk}},\ }\href@noop {} {\bibfield  {journal} {\bibinfo  {journal} {Accounts of chemical research}\ }\textbf {\bibinfo {volume} {48}},\ \bibinfo {pages} {1938} (\bibinfo {year} {2015})}\BibitemShut {NoStop}%
\bibitem [{\citenamefont {Palacci}\ \emph {et~al.}(2014)\citenamefont {Palacci}, \citenamefont {Sacanna}, \citenamefont {Kim}, \citenamefont {Yi}, \citenamefont {Pine},\ and\ \citenamefont {Chaikin}}]{palacci2014light}%
  \BibitemOpen
  \bibfield  {author} {\bibinfo {author} {\bibfnamefont {J.}~\bibnamefont {Palacci}}, \bibinfo {author} {\bibfnamefont {S.}~\bibnamefont {Sacanna}}, \bibinfo {author} {\bibfnamefont {S.-H.}\ \bibnamefont {Kim}}, \bibinfo {author} {\bibfnamefont {G.-R.}\ \bibnamefont {Yi}}, \bibinfo {author} {\bibfnamefont {D.~J.}\ \bibnamefont {Pine}}, \ and\ \bibinfo {author} {\bibfnamefont {P.~M.}\ \bibnamefont {Chaikin}},\ }\href@noop {} {\bibfield  {journal} {\bibinfo  {journal} {Philosophical Transactions of the Royal Society A: Mathematical, Physical and Engineering Sciences}\ }\textbf {\bibinfo {volume} {372}},\ \bibinfo {pages} {20130372} (\bibinfo {year} {2014})}\BibitemShut {NoStop}%
\bibitem [{\citenamefont {Buttinoni}\ \emph {et~al.}(2013)\citenamefont {Buttinoni}, \citenamefont {Bialk{\'e}}, \citenamefont {K{\"u}mmel}, \citenamefont {L{\"o}wen}, \citenamefont {Bechinger},\ and\ \citenamefont {Speck}}]{buttinoni2013dynamical}%
  \BibitemOpen
  \bibfield  {author} {\bibinfo {author} {\bibfnamefont {I.}~\bibnamefont {Buttinoni}}, \bibinfo {author} {\bibfnamefont {J.}~\bibnamefont {Bialk{\'e}}}, \bibinfo {author} {\bibfnamefont {F.}~\bibnamefont {K{\"u}mmel}}, \bibinfo {author} {\bibfnamefont {H.}~\bibnamefont {L{\"o}wen}}, \bibinfo {author} {\bibfnamefont {C.}~\bibnamefont {Bechinger}}, \ and\ \bibinfo {author} {\bibfnamefont {T.}~\bibnamefont {Speck}},\ }\href@noop {} {\bibfield  {journal} {\bibinfo  {journal} {Physical review letters}\ }\textbf {\bibinfo {volume} {110}},\ \bibinfo {pages} {238301} (\bibinfo {year} {2013})}\BibitemShut {NoStop}%
\bibitem [{\citenamefont {Brambilla}\ \emph {et~al.}(2013)\citenamefont {Brambilla}, \citenamefont {Ferrante}, \citenamefont {Birattari},\ and\ \citenamefont {Dorigo}}]{brambilla2013swarm}%
  \BibitemOpen
  \bibfield  {author} {\bibinfo {author} {\bibfnamefont {M.}~\bibnamefont {Brambilla}}, \bibinfo {author} {\bibfnamefont {E.}~\bibnamefont {Ferrante}}, \bibinfo {author} {\bibfnamefont {M.}~\bibnamefont {Birattari}}, \ and\ \bibinfo {author} {\bibfnamefont {M.}~\bibnamefont {Dorigo}},\ }\href@noop {} {\bibfield  {journal} {\bibinfo  {journal} {Swarm Intelligence}\ }\textbf {\bibinfo {volume} {7}},\ \bibinfo {pages} {1} (\bibinfo {year} {2013})}\BibitemShut {NoStop}%
\bibitem [{\citenamefont {Vicsek}\ \emph {et~al.}(1995)\citenamefont {Vicsek}, \citenamefont {Czir{\'o}k}, \citenamefont {Ben-Jacob}, \citenamefont {Cohen},\ and\ \citenamefont {Shochet}}]{vicsek1995novel}%
  \BibitemOpen
  \bibfield  {author} {\bibinfo {author} {\bibfnamefont {T.}~\bibnamefont {Vicsek}}, \bibinfo {author} {\bibfnamefont {A.}~\bibnamefont {Czir{\'o}k}}, \bibinfo {author} {\bibfnamefont {E.}~\bibnamefont {Ben-Jacob}}, \bibinfo {author} {\bibfnamefont {I.}~\bibnamefont {Cohen}}, \ and\ \bibinfo {author} {\bibfnamefont {O.}~\bibnamefont {Shochet}},\ }\href@noop {} {\bibfield  {journal} {\bibinfo  {journal} {Physical review letters}\ }\textbf {\bibinfo {volume} {75}},\ \bibinfo {pages} {1226} (\bibinfo {year} {1995})}\BibitemShut {NoStop}%
\bibitem [{\citenamefont {Vicsek}(2012)}]{vicsek2012collective}%
  \BibitemOpen
  \bibfield  {author} {\bibinfo {author} {\bibfnamefont {T.}~\bibnamefont {Vicsek}},\ }\href@noop {} {\enquote {\bibinfo {title} {Collective motion},}\ } (\bibinfo {year} {2012})\BibitemShut {NoStop}%
\bibitem [{\citenamefont {Das}\ \emph {et~al.}(2024)\citenamefont {Das}, \citenamefont {Ciarchi}, \citenamefont {Zhou}, \citenamefont {Yan}, \citenamefont {Zhang},\ and\ \citenamefont {Alert}}]{das2024flocking}%
  \BibitemOpen
  \bibfield  {author} {\bibinfo {author} {\bibfnamefont {S.}~\bibnamefont {Das}}, \bibinfo {author} {\bibfnamefont {M.}~\bibnamefont {Ciarchi}}, \bibinfo {author} {\bibfnamefont {Z.}~\bibnamefont {Zhou}}, \bibinfo {author} {\bibfnamefont {J.}~\bibnamefont {Yan}}, \bibinfo {author} {\bibfnamefont {J.}~\bibnamefont {Zhang}}, \ and\ \bibinfo {author} {\bibfnamefont {R.}~\bibnamefont {Alert}},\ }\href@noop {} {\bibfield  {journal} {\bibinfo  {journal} {Physical Review X}\ }\textbf {\bibinfo {volume} {14}},\ \bibinfo {pages} {031008} (\bibinfo {year} {2024})}\BibitemShut {NoStop}%
\bibitem [{\citenamefont {Chat{\'e}}\ \emph {et~al.}(2008)\citenamefont {Chat{\'e}}, \citenamefont {Ginelli}, \citenamefont {Gr{\'e}goire},\ and\ \citenamefont {Raynaud}}]{chate2008collective}%
  \BibitemOpen
  \bibfield  {author} {\bibinfo {author} {\bibfnamefont {H.}~\bibnamefont {Chat{\'e}}}, \bibinfo {author} {\bibfnamefont {F.}~\bibnamefont {Ginelli}}, \bibinfo {author} {\bibfnamefont {G.}~\bibnamefont {Gr{\'e}goire}}, \ and\ \bibinfo {author} {\bibfnamefont {F.}~\bibnamefont {Raynaud}},\ }\href@noop {} {\bibfield  {journal} {\bibinfo  {journal} {Physical Review E—Statistical, Nonlinear, and Soft Matter Physics}\ }\textbf {\bibinfo {volume} {77}},\ \bibinfo {pages} {046113} (\bibinfo {year} {2008})}\BibitemShut {NoStop}%
\bibitem [{\citenamefont {O'Loan}\ and\ \citenamefont {Evans}(1999)}]{o1999alternating}%
  \BibitemOpen
  \bibfield  {author} {\bibinfo {author} {\bibfnamefont {O.}~\bibnamefont {O'Loan}}\ and\ \bibinfo {author} {\bibfnamefont {M.}~\bibnamefont {Evans}},\ }\href@noop {} {\bibfield  {journal} {\bibinfo  {journal} {Journal of Physics A: Mathematical and General}\ }\textbf {\bibinfo {volume} {32}},\ \bibinfo {pages} {L99} (\bibinfo {year} {1999})}\BibitemShut {NoStop}%
\bibitem [{\citenamefont {Czir{\'o}k}\ \emph {et~al.}(1999)\citenamefont {Czir{\'o}k}, \citenamefont {Barab{\'a}si},\ and\ \citenamefont {Vicsek}}]{czirok1999collective}%
  \BibitemOpen
  \bibfield  {author} {\bibinfo {author} {\bibfnamefont {A.}~\bibnamefont {Czir{\'o}k}}, \bibinfo {author} {\bibfnamefont {A.-L.}\ \bibnamefont {Barab{\'a}si}}, \ and\ \bibinfo {author} {\bibfnamefont {T.}~\bibnamefont {Vicsek}},\ }\href@noop {} {\bibfield  {journal} {\bibinfo  {journal} {Physical Review Letters}\ }\textbf {\bibinfo {volume} {82}},\ \bibinfo {pages} {209} (\bibinfo {year} {1999})}\BibitemShut {NoStop}%
\bibitem [{\citenamefont {Raymond}\ and\ \citenamefont {Evans}(2006)}]{raymond2006flocking}%
  \BibitemOpen
  \bibfield  {author} {\bibinfo {author} {\bibfnamefont {J.}~\bibnamefont {Raymond}}\ and\ \bibinfo {author} {\bibfnamefont {M.}~\bibnamefont {Evans}},\ }\href@noop {} {\bibfield  {journal} {\bibinfo  {journal} {Physical Review E}\ }\textbf {\bibinfo {volume} {73}},\ \bibinfo {pages} {036112} (\bibinfo {year} {2006})}\BibitemShut {NoStop}%
\bibitem [{\citenamefont {Buhl}\ \emph {et~al.}(2006)\citenamefont {Buhl}, \citenamefont {Sumpter}, \citenamefont {Couzin}, \citenamefont {Hale}, \citenamefont {Despland}, \citenamefont {Miller},\ and\ \citenamefont {Simpson}}]{buhl2006disorder}%
  \BibitemOpen
  \bibfield  {author} {\bibinfo {author} {\bibfnamefont {J.}~\bibnamefont {Buhl}}, \bibinfo {author} {\bibfnamefont {D.~J.}\ \bibnamefont {Sumpter}}, \bibinfo {author} {\bibfnamefont {I.~D.}\ \bibnamefont {Couzin}}, \bibinfo {author} {\bibfnamefont {J.~J.}\ \bibnamefont {Hale}}, \bibinfo {author} {\bibfnamefont {E.}~\bibnamefont {Despland}}, \bibinfo {author} {\bibfnamefont {E.~R.}\ \bibnamefont {Miller}}, \ and\ \bibinfo {author} {\bibfnamefont {S.~J.}\ \bibnamefont {Simpson}},\ }\href@noop {} {\bibfield  {journal} {\bibinfo  {journal} {Science}\ }\textbf {\bibinfo {volume} {312}},\ \bibinfo {pages} {1402} (\bibinfo {year} {2006})}\BibitemShut {NoStop}%
\bibitem [{\citenamefont {Yates}\ \emph {et~al.}(2009)\citenamefont {Yates}, \citenamefont {Erban}, \citenamefont {Escudero}, \citenamefont {Couzin}, \citenamefont {Buhl}, \citenamefont {Kevrekidis}, \citenamefont {Maini},\ and\ \citenamefont {Sumpter}}]{yates2009inherent}%
  \BibitemOpen
  \bibfield  {author} {\bibinfo {author} {\bibfnamefont {C.~A.}\ \bibnamefont {Yates}}, \bibinfo {author} {\bibfnamefont {R.}~\bibnamefont {Erban}}, \bibinfo {author} {\bibfnamefont {C.}~\bibnamefont {Escudero}}, \bibinfo {author} {\bibfnamefont {I.~D.}\ \bibnamefont {Couzin}}, \bibinfo {author} {\bibfnamefont {J.}~\bibnamefont {Buhl}}, \bibinfo {author} {\bibfnamefont {I.~G.}\ \bibnamefont {Kevrekidis}}, \bibinfo {author} {\bibfnamefont {P.~K.}\ \bibnamefont {Maini}}, \ and\ \bibinfo {author} {\bibfnamefont {D.~J.}\ \bibnamefont {Sumpter}},\ }\href@noop {} {\bibfield  {journal} {\bibinfo  {journal} {Proceedings of the National Academy of Sciences}\ }\textbf {\bibinfo {volume} {106}},\ \bibinfo {pages} {5464} (\bibinfo {year} {2009})}\BibitemShut {NoStop}%
\bibitem [{\citenamefont {Bode}\ \emph {et~al.}(2010)\citenamefont {Bode}, \citenamefont {Franks},\ and\ \citenamefont {Wood}}]{bode2010making}%
  \BibitemOpen
  \bibfield  {author} {\bibinfo {author} {\bibfnamefont {N.~W.}\ \bibnamefont {Bode}}, \bibinfo {author} {\bibfnamefont {D.~W.}\ \bibnamefont {Franks}}, \ and\ \bibinfo {author} {\bibfnamefont {A.~J.}\ \bibnamefont {Wood}},\ }\href@noop {} {\bibfield  {journal} {\bibinfo  {journal} {Journal of theoretical biology}\ }\textbf {\bibinfo {volume} {267}},\ \bibinfo {pages} {292} (\bibinfo {year} {2010})}\BibitemShut {NoStop}%
\bibitem [{\citenamefont {Dossetti}(2011)}]{dossetti2011cohesive}%
  \BibitemOpen
  \bibfield  {author} {\bibinfo {author} {\bibfnamefont {V.}~\bibnamefont {Dossetti}},\ }\href@noop {} {\bibfield  {journal} {\bibinfo  {journal} {Journal of Physics A: Mathematical and Theoretical}\ }\textbf {\bibinfo {volume} {45}},\ \bibinfo {pages} {035003} (\bibinfo {year} {2011})}\BibitemShut {NoStop}%
\bibitem [{\citenamefont {Solon}\ and\ \citenamefont {Tailleur}(2013)}]{solon2013revisiting}%
  \BibitemOpen
  \bibfield  {author} {\bibinfo {author} {\bibfnamefont {A.~P.}\ \bibnamefont {Solon}}\ and\ \bibinfo {author} {\bibfnamefont {J.}~\bibnamefont {Tailleur}},\ }\href@noop {} {\bibfield  {journal} {\bibinfo  {journal} {Physical review letters}\ }\textbf {\bibinfo {volume} {111}},\ \bibinfo {pages} {078101} (\bibinfo {year} {2013})}\BibitemShut {NoStop}%
\bibitem [{\citenamefont {Sakaguchi}\ and\ \citenamefont {Ishibashi}(2019)}]{sakaguchi2019flip}%
  \BibitemOpen
  \bibfield  {author} {\bibinfo {author} {\bibfnamefont {H.}~\bibnamefont {Sakaguchi}}\ and\ \bibinfo {author} {\bibfnamefont {K.}~\bibnamefont {Ishibashi}},\ }\href@noop {} {\bibfield  {journal} {\bibinfo  {journal} {Physical Review E}\ }\textbf {\bibinfo {volume} {100}},\ \bibinfo {pages} {052113} (\bibinfo {year} {2019})}\BibitemShut {NoStop}%
\bibitem [{\citenamefont {Benvegnen}\ \emph {et~al.}(2022)\citenamefont {Benvegnen}, \citenamefont {Chat{\'e}}, \citenamefont {Krapivsky}, \citenamefont {Tailleur},\ and\ \citenamefont {Solon}}]{benvegnen2022flocking}%
  \BibitemOpen
  \bibfield  {author} {\bibinfo {author} {\bibfnamefont {B.}~\bibnamefont {Benvegnen}}, \bibinfo {author} {\bibfnamefont {H.}~\bibnamefont {Chat{\'e}}}, \bibinfo {author} {\bibfnamefont {P.~L.}\ \bibnamefont {Krapivsky}}, \bibinfo {author} {\bibfnamefont {J.}~\bibnamefont {Tailleur}}, \ and\ \bibinfo {author} {\bibfnamefont {A.}~\bibnamefont {Solon}},\ }\href@noop {} {\bibfield  {journal} {\bibinfo  {journal} {Physical Review E}\ }\textbf {\bibinfo {volume} {106}},\ \bibinfo {pages} {054608} (\bibinfo {year} {2022})}\BibitemShut {NoStop}%
\bibitem [{\citenamefont {Kumar}\ \emph {et~al.}(2024)\citenamefont {Kumar}, \citenamefont {Pattanayak}, \citenamefont {Singh},\ and\ \citenamefont {Mishra}}]{KUMAR2024129773}%
  \BibitemOpen
  \bibfield  {author} {\bibinfo {author} {\bibfnamefont {A.}~\bibnamefont {Kumar}}, \bibinfo {author} {\bibfnamefont {S.}~\bibnamefont {Pattanayak}}, \bibinfo {author} {\bibfnamefont {R.}~\bibnamefont {Singh}}, \ and\ \bibinfo {author} {\bibfnamefont {S.}~\bibnamefont {Mishra}},\ }\href@noop {} {\bibfield  {journal} {\bibinfo  {journal} {Physics Letters A}\ }\textbf {\bibinfo {volume} {523}},\ \bibinfo {pages} {129773} (\bibinfo {year} {2024})}\BibitemShut {NoStop}%
\bibitem [{\citenamefont {Mishra}\ \emph {et~al.}(2024)\citenamefont {Mishra}, \citenamefont {Puitandy},\ and\ \citenamefont {Mishra}}]{pawanDirectionalCue}%
  \BibitemOpen
  \bibfield  {author} {\bibinfo {author} {\bibfnamefont {P.~K.}\ \bibnamefont {Mishra}}, \bibinfo {author} {\bibfnamefont {A.}~\bibnamefont {Puitandy}}, \ and\ \bibinfo {author} {\bibfnamefont {S.}~\bibnamefont {Mishra}},\ }\href@noop {} {\bibfield  {journal} {\bibinfo  {journal} {Europhysics Letters}\ } (\bibinfo {year} {2024})}\BibitemShut {NoStop}%
\bibitem [{\citenamefont {Laighl{\'e}is}\ \emph {et~al.}(2018)\citenamefont {Laighl{\'e}is}, \citenamefont {Evans},\ and\ \citenamefont {Blythe}}]{laighleis2018minimal}%
  \BibitemOpen
  \bibfield  {author} {\bibinfo {author} {\bibfnamefont {E.~{\'O}.}\ \bibnamefont {Laighl{\'e}is}}, \bibinfo {author} {\bibfnamefont {M.~R.}\ \bibnamefont {Evans}}, \ and\ \bibinfo {author} {\bibfnamefont {R.~A.}\ \bibnamefont {Blythe}},\ }\href@noop {} {\bibfield  {journal} {\bibinfo  {journal} {Physical Review E}\ }\textbf {\bibinfo {volume} {98}},\ \bibinfo {pages} {062127} (\bibinfo {year} {2018})}\BibitemShut {NoStop}%
\bibitem [{\citenamefont {Durve}\ \emph {et~al.}(2020)\citenamefont {Durve}, \citenamefont {Peruani},\ and\ \citenamefont {Celani}}]{durve2020learning}%
  \BibitemOpen
  \bibfield  {author} {\bibinfo {author} {\bibfnamefont {M.}~\bibnamefont {Durve}}, \bibinfo {author} {\bibfnamefont {F.}~\bibnamefont {Peruani}}, \ and\ \bibinfo {author} {\bibfnamefont {A.}~\bibnamefont {Celani}},\ }\href@noop {} {\bibfield  {journal} {\bibinfo  {journal} {Physical Review E}\ }\textbf {\bibinfo {volume} {102}},\ \bibinfo {pages} {012601} (\bibinfo {year} {2020})}\BibitemShut {NoStop}%
\bibitem [{\citenamefont {Khlif}\ \emph {et~al.}(2022)\citenamefont {Khlif}, \citenamefont {Khraief},\ and\ \citenamefont {Belghith}}]{khlif2022reinforcement}%
  \BibitemOpen
  \bibfield  {author} {\bibinfo {author} {\bibfnamefont {N.}~\bibnamefont {Khlif}}, \bibinfo {author} {\bibfnamefont {N.}~\bibnamefont {Khraief}}, \ and\ \bibinfo {author} {\bibfnamefont {S.}~\bibnamefont {Belghith}},\ }\href@noop {} {\bibfield  {journal} {\bibinfo  {journal} {2022 IEEE Information Technologies \& Smart Industrial Systems (ITSIS)}\ ,\ \bibinfo {pages} {1}} (\bibinfo {year} {2022})}\BibitemShut {NoStop}%
\bibitem [{\citenamefont {Pan}\ \emph {et~al.}(2022)\citenamefont {Pan}, \citenamefont {Xiang}, \citenamefont {Wang}, \citenamefont {Yu},\ and\ \citenamefont {Zhou}}]{pan2022research}%
  \BibitemOpen
  \bibfield  {author} {\bibinfo {author} {\bibfnamefont {G.}~\bibnamefont {Pan}}, \bibinfo {author} {\bibfnamefont {Y.}~\bibnamefont {Xiang}}, \bibinfo {author} {\bibfnamefont {X.}~\bibnamefont {Wang}}, \bibinfo {author} {\bibfnamefont {Z.}~\bibnamefont {Yu}}, \ and\ \bibinfo {author} {\bibfnamefont {X.}~\bibnamefont {Zhou}},\ }\href@noop {} {\bibfield  {journal} {\bibinfo  {journal} {Soft Computing}\ }\textbf {\bibinfo {volume} {26}},\ \bibinfo {pages} {8961} (\bibinfo {year} {2022})}\BibitemShut {NoStop}%
\bibitem [{\citenamefont {Gharbi}(2024)}]{gharbi2024dynamic}%
  \BibitemOpen
  \bibfield  {author} {\bibinfo {author} {\bibfnamefont {A.}~\bibnamefont {Gharbi}},\ }\href@noop {} {\bibfield  {journal} {\bibinfo  {journal} {Applied Computing and Informatics}\ } (\bibinfo {year} {2024})}\BibitemShut {NoStop}%
\bibitem [{\citenamefont {Colabrese}\ \emph {et~al.}(2017)\citenamefont {Colabrese}, \citenamefont {Gustavsson}, \citenamefont {Celani},\ and\ \citenamefont {Biferale}}]{colabrese2017flow}%
  \BibitemOpen
  \bibfield  {author} {\bibinfo {author} {\bibfnamefont {S.}~\bibnamefont {Colabrese}}, \bibinfo {author} {\bibfnamefont {K.}~\bibnamefont {Gustavsson}}, \bibinfo {author} {\bibfnamefont {A.}~\bibnamefont {Celani}}, \ and\ \bibinfo {author} {\bibfnamefont {L.}~\bibnamefont {Biferale}},\ }\href@noop {} {\bibfield  {journal} {\bibinfo  {journal} {Physical review letters}\ }\textbf {\bibinfo {volume} {118}},\ \bibinfo {pages} {158004} (\bibinfo {year} {2017})}\BibitemShut {NoStop}%
\bibitem [{\citenamefont {Nasiri}\ \emph {et~al.}(2024)\citenamefont {Nasiri}, \citenamefont {Loran},\ and\ \citenamefont {Liebchen}}]{nasiri2024smart}%
  \BibitemOpen
  \bibfield  {author} {\bibinfo {author} {\bibfnamefont {M.}~\bibnamefont {Nasiri}}, \bibinfo {author} {\bibfnamefont {E.}~\bibnamefont {Loran}}, \ and\ \bibinfo {author} {\bibfnamefont {B.}~\bibnamefont {Liebchen}},\ }\href@noop {} {\bibfield  {journal} {\bibinfo  {journal} {Proceedings of the National Academy of Sciences}\ }\textbf {\bibinfo {volume} {121}},\ \bibinfo {pages} {e2317618121} (\bibinfo {year} {2024})}\BibitemShut {NoStop}%
\bibitem [{\citenamefont {Schneider}\ and\ \citenamefont {Stark}(2019)}]{schneider2019optimal}%
  \BibitemOpen
  \bibfield  {author} {\bibinfo {author} {\bibfnamefont {E.}~\bibnamefont {Schneider}}\ and\ \bibinfo {author} {\bibfnamefont {H.}~\bibnamefont {Stark}},\ }\href@noop {} {\bibfield  {journal} {\bibinfo  {journal} {Europhysics Letters}\ }\textbf {\bibinfo {volume} {127}},\ \bibinfo {pages} {64003} (\bibinfo {year} {2019})}\BibitemShut {NoStop}%
\bibitem [{\citenamefont {Putzke}\ and\ \citenamefont {Stark}(2023)}]{putzke2023optimal}%
  \BibitemOpen
  \bibfield  {author} {\bibinfo {author} {\bibfnamefont {M.}~\bibnamefont {Putzke}}\ and\ \bibinfo {author} {\bibfnamefont {H.}~\bibnamefont {Stark}},\ }\href@noop {} {\bibfield  {journal} {\bibinfo  {journal} {The European Physical Journal E}\ }\textbf {\bibinfo {volume} {46}},\ \bibinfo {pages} {48} (\bibinfo {year} {2023})}\BibitemShut {NoStop}%
\bibitem [{\citenamefont {Nasiri}\ \emph {et~al.}(2023)\citenamefont {Nasiri}, \citenamefont {L{\"o}wen},\ and\ \citenamefont {Liebchen}}]{nasiri2023optimal}%
  \BibitemOpen
  \bibfield  {author} {\bibinfo {author} {\bibfnamefont {M.}~\bibnamefont {Nasiri}}, \bibinfo {author} {\bibfnamefont {H.}~\bibnamefont {L{\"o}wen}}, \ and\ \bibinfo {author} {\bibfnamefont {B.}~\bibnamefont {Liebchen}},\ }\href@noop {} {\bibfield  {journal} {\bibinfo  {journal} {Europhysics Letters}\ }\textbf {\bibinfo {volume} {142}},\ \bibinfo {pages} {17001} (\bibinfo {year} {2023})}\BibitemShut {NoStop}%
\bibitem [{\citenamefont {Nasiri}\ and\ \citenamefont {Liebchen}(2022)}]{nasiri2022reinforcement}%
  \BibitemOpen
  \bibfield  {author} {\bibinfo {author} {\bibfnamefont {M.}~\bibnamefont {Nasiri}}\ and\ \bibinfo {author} {\bibfnamefont {B.}~\bibnamefont {Liebchen}},\ }\href@noop {} {\bibfield  {journal} {\bibinfo  {journal} {New Journal of Physics}\ }\textbf {\bibinfo {volume} {24}},\ \bibinfo {pages} {073042} (\bibinfo {year} {2022})}\BibitemShut {NoStop}%
\bibitem [{\citenamefont {Monderkamp}\ \emph {et~al.}(2022)\citenamefont {Monderkamp}, \citenamefont {Schwarzendahl}, \citenamefont {Klatt},\ and\ \citenamefont {L{\"o}wen}}]{monderkamp2022active}%
  \BibitemOpen
  \bibfield  {author} {\bibinfo {author} {\bibfnamefont {P.~A.}\ \bibnamefont {Monderkamp}}, \bibinfo {author} {\bibfnamefont {F.~J.}\ \bibnamefont {Schwarzendahl}}, \bibinfo {author} {\bibfnamefont {M.~A.}\ \bibnamefont {Klatt}}, \ and\ \bibinfo {author} {\bibfnamefont {H.}~\bibnamefont {L{\"o}wen}},\ }\href@noop {} {\bibfield  {journal} {\bibinfo  {journal} {Machine Learning: Science and Technology}\ }\textbf {\bibinfo {volume} {3}},\ \bibinfo {pages} {045024} (\bibinfo {year} {2022})}\BibitemShut {NoStop}%
\bibitem [{\citenamefont {Biferale}\ \emph {et~al.}(2019)\citenamefont {Biferale}, \citenamefont {Bonaccorso}, \citenamefont {Buzzicotti}, \citenamefont {Clark Di~Leoni},\ and\ \citenamefont {Gustavsson}}]{biferale2019zermelo}%
  \BibitemOpen
  \bibfield  {author} {\bibinfo {author} {\bibfnamefont {L.}~\bibnamefont {Biferale}}, \bibinfo {author} {\bibfnamefont {F.}~\bibnamefont {Bonaccorso}}, \bibinfo {author} {\bibfnamefont {M.}~\bibnamefont {Buzzicotti}}, \bibinfo {author} {\bibfnamefont {P.}~\bibnamefont {Clark Di~Leoni}}, \ and\ \bibinfo {author} {\bibfnamefont {K.}~\bibnamefont {Gustavsson}},\ }\href@noop {} {\bibfield  {journal} {\bibinfo  {journal} {Chaos: An Interdisciplinary Journal of Nonlinear Science}\ }\textbf {\bibinfo {volume} {29}} (\bibinfo {year} {2019})}\BibitemShut {NoStop}%
\bibitem [{\citenamefont {Alageshan}\ \emph {et~al.}(2020)\citenamefont {Alageshan}, \citenamefont {Verma}, \citenamefont {Bec},\ and\ \citenamefont {Pandit}}]{alageshan2020machine}%
  \BibitemOpen
  \bibfield  {author} {\bibinfo {author} {\bibfnamefont {J.~K.}\ \bibnamefont {Alageshan}}, \bibinfo {author} {\bibfnamefont {A.~K.}\ \bibnamefont {Verma}}, \bibinfo {author} {\bibfnamefont {J.}~\bibnamefont {Bec}}, \ and\ \bibinfo {author} {\bibfnamefont {R.}~\bibnamefont {Pandit}},\ }\href@noop {} {\bibfield  {journal} {\bibinfo  {journal} {Physical Review E}\ }\textbf {\bibinfo {volume} {101}},\ \bibinfo {pages} {043110} (\bibinfo {year} {2020})}\BibitemShut {NoStop}%
\bibitem [{\citenamefont {Buzzicotti}\ \emph {et~al.}(2020)\citenamefont {Buzzicotti}, \citenamefont {Biferale}, \citenamefont {Bonaccorso}, \citenamefont {Clark~di Leoni},\ and\ \citenamefont {Gustavsson}}]{buzzicotti2020optimal}%
  \BibitemOpen
  \bibfield  {author} {\bibinfo {author} {\bibfnamefont {M.}~\bibnamefont {Buzzicotti}}, \bibinfo {author} {\bibfnamefont {L.}~\bibnamefont {Biferale}}, \bibinfo {author} {\bibfnamefont {F.}~\bibnamefont {Bonaccorso}}, \bibinfo {author} {\bibfnamefont {P.}~\bibnamefont {Clark~di Leoni}}, \ and\ \bibinfo {author} {\bibfnamefont {K.}~\bibnamefont {Gustavsson}},\ }in\ \href@noop {} {\emph {\bibinfo {booktitle} {International Conference of the Italian Association for Artificial Intelligence}}}\ (\bibinfo {organization} {Springer},\ \bibinfo {year} {2020})\ pp.\ \bibinfo {pages} {223--234}\BibitemShut {NoStop}%
\bibitem [{\citenamefont {Zou}\ \emph {et~al.}(2022)\citenamefont {Zou}, \citenamefont {Liu}, \citenamefont {Young}, \citenamefont {Pak},\ and\ \citenamefont {Tsang}}]{zou2022gait}%
  \BibitemOpen
  \bibfield  {author} {\bibinfo {author} {\bibfnamefont {Z.}~\bibnamefont {Zou}}, \bibinfo {author} {\bibfnamefont {Y.}~\bibnamefont {Liu}}, \bibinfo {author} {\bibfnamefont {Y.-N.}\ \bibnamefont {Young}}, \bibinfo {author} {\bibfnamefont {O.~S.}\ \bibnamefont {Pak}}, \ and\ \bibinfo {author} {\bibfnamefont {A.~C.}\ \bibnamefont {Tsang}},\ }\href@noop {} {\bibfield  {journal} {\bibinfo  {journal} {Communications Physics}\ }\textbf {\bibinfo {volume} {5}},\ \bibinfo {pages} {158} (\bibinfo {year} {2022})}\BibitemShut {NoStop}%
\bibitem [{\citenamefont {Falk}\ \emph {et~al.}(2021)\citenamefont {Falk}, \citenamefont {Alizadehyazdi}, \citenamefont {Jaeger},\ and\ \citenamefont {Murugan}}]{falk2021learning}%
  \BibitemOpen
  \bibfield  {author} {\bibinfo {author} {\bibfnamefont {M.~J.}\ \bibnamefont {Falk}}, \bibinfo {author} {\bibfnamefont {V.}~\bibnamefont {Alizadehyazdi}}, \bibinfo {author} {\bibfnamefont {H.}~\bibnamefont {Jaeger}}, \ and\ \bibinfo {author} {\bibfnamefont {A.}~\bibnamefont {Murugan}},\ }\href@noop {} {\bibfield  {journal} {\bibinfo  {journal} {Physical Review Research}\ }\textbf {\bibinfo {volume} {3}},\ \bibinfo {pages} {033291} (\bibinfo {year} {2021})}\BibitemShut {NoStop}%
\bibitem [{\citenamefont {Gerhard}\ \emph {et~al.}(2021)\citenamefont {Gerhard}, \citenamefont {Jayaram}, \citenamefont {Fischer},\ and\ \citenamefont {Speck}}]{gerhard2021hunting}%
  \BibitemOpen
  \bibfield  {author} {\bibinfo {author} {\bibfnamefont {M.}~\bibnamefont {Gerhard}}, \bibinfo {author} {\bibfnamefont {A.}~\bibnamefont {Jayaram}}, \bibinfo {author} {\bibfnamefont {A.}~\bibnamefont {Fischer}}, \ and\ \bibinfo {author} {\bibfnamefont {T.}~\bibnamefont {Speck}},\ }\href@noop {} {\bibfield  {journal} {\bibinfo  {journal} {Physical Review E}\ }\textbf {\bibinfo {volume} {104}},\ \bibinfo {pages} {054614} (\bibinfo {year} {2021})}\BibitemShut {NoStop}%
\bibitem [{\citenamefont {Yang}\ \emph {et~al.}(2020)\citenamefont {Yang}, \citenamefont {Bevan},\ and\ \citenamefont {Li}}]{yang2020micro}%
  \BibitemOpen
  \bibfield  {author} {\bibinfo {author} {\bibfnamefont {Y.}~\bibnamefont {Yang}}, \bibinfo {author} {\bibfnamefont {M.~A.}\ \bibnamefont {Bevan}}, \ and\ \bibinfo {author} {\bibfnamefont {B.}~\bibnamefont {Li}},\ }\href@noop {} {\bibfield  {journal} {\bibinfo  {journal} {Advanced Theory and Simulations}\ }\textbf {\bibinfo {volume} {3}},\ \bibinfo {pages} {2000034} (\bibinfo {year} {2020})}\BibitemShut {NoStop}%
\bibitem [{\citenamefont {Mui{\~n}os-Landin}\ \emph {et~al.}(2021)\citenamefont {Mui{\~n}os-Landin}, \citenamefont {Fischer}, \citenamefont {Holubec},\ and\ \citenamefont {Cichos}}]{muinos2021reinforcement}%
  \BibitemOpen
  \bibfield  {author} {\bibinfo {author} {\bibfnamefont {S.}~\bibnamefont {Mui{\~n}os-Landin}}, \bibinfo {author} {\bibfnamefont {A.}~\bibnamefont {Fischer}}, \bibinfo {author} {\bibfnamefont {V.}~\bibnamefont {Holubec}}, \ and\ \bibinfo {author} {\bibfnamefont {F.}~\bibnamefont {Cichos}},\ }\href@noop {} {\bibfield  {journal} {\bibinfo  {journal} {Science Robotics}\ }\textbf {\bibinfo {volume} {6}},\ \bibinfo {pages} {eabd9285} (\bibinfo {year} {2021})}\BibitemShut {NoStop}%
\bibitem [{\citenamefont {Pattanayak}\ and\ \citenamefont {Mishra}(2018)}]{pattanayak2018collection}%
  \BibitemOpen
  \bibfield  {author} {\bibinfo {author} {\bibfnamefont {S.}~\bibnamefont {Pattanayak}}\ and\ \bibinfo {author} {\bibfnamefont {S.}~\bibnamefont {Mishra}},\ }\href@noop {} {\bibfield  {journal} {\bibinfo  {journal} {Journal of Physics Communications}\ }\textbf {\bibinfo {volume} {2}},\ \bibinfo {pages} {045007} (\bibinfo {year} {2018})}\BibitemShut {NoStop}%
\bibitem [{\citenamefont {Singh}\ \emph {et~al.}(2021{\natexlab{a}})\citenamefont {Singh}, \citenamefont {Pattanayak},\ and\ \citenamefont {Mishra}}]{singh2021ordering}%
  \BibitemOpen
  \bibfield  {author} {\bibinfo {author} {\bibfnamefont {J.~P.}\ \bibnamefont {Singh}}, \bibinfo {author} {\bibfnamefont {S.}~\bibnamefont {Pattanayak}}, \ and\ \bibinfo {author} {\bibfnamefont {S.}~\bibnamefont {Mishra}},\ }\href@noop {} {\bibfield  {journal} {\bibinfo  {journal} {Journal of Physics A: Mathematical and Theoretical}\ }\textbf {\bibinfo {volume} {54}},\ \bibinfo {pages} {115001} (\bibinfo {year} {2021}{\natexlab{a}})}\BibitemShut {NoStop}%
\bibitem [{\citenamefont {Singh}\ \emph {et~al.}(2021{\natexlab{b}})\citenamefont {Singh}, \citenamefont {Kumar},\ and\ \citenamefont {Mishra}}]{singh2021bond}%
  \BibitemOpen
  \bibfield  {author} {\bibinfo {author} {\bibfnamefont {J.~P.}\ \bibnamefont {Singh}}, \bibinfo {author} {\bibfnamefont {S.}~\bibnamefont {Kumar}}, \ and\ \bibinfo {author} {\bibfnamefont {S.}~\bibnamefont {Mishra}},\ }\href@noop {} {\bibfield  {journal} {\bibinfo  {journal} {Journal of Statistical Mechanics: Theory and Experiment}\ }\textbf {\bibinfo {volume} {2021}},\ \bibinfo {pages} {083217} (\bibinfo {year} {2021}{\natexlab{b}})}\BibitemShut {NoStop}%
\bibitem [{\citenamefont {Katz}\ \emph {et~al.}(2011)\citenamefont {Katz}, \citenamefont {Tunstr{\o}m}, \citenamefont {Ioannou}, \citenamefont {Huepe},\ and\ \citenamefont {Couzin}}]{katz2011inferring}%
  \BibitemOpen
  \bibfield  {author} {\bibinfo {author} {\bibfnamefont {Y.}~\bibnamefont {Katz}}, \bibinfo {author} {\bibfnamefont {K.}~\bibnamefont {Tunstr{\o}m}}, \bibinfo {author} {\bibfnamefont {C.~C.}\ \bibnamefont {Ioannou}}, \bibinfo {author} {\bibfnamefont {C.}~\bibnamefont {Huepe}}, \ and\ \bibinfo {author} {\bibfnamefont {I.~D.}\ \bibnamefont {Couzin}},\ }\href@noop {} {\bibfield  {journal} {\bibinfo  {journal} {Proceedings of the National Academy of Sciences}\ }\textbf {\bibinfo {volume} {108}},\ \bibinfo {pages} {18720} (\bibinfo {year} {2011})}\BibitemShut {NoStop}%
\bibitem [{\citenamefont {Cisneros}\ \emph {et~al.}(2011)\citenamefont {Cisneros}, \citenamefont {Kessler}, \citenamefont {Ganguly},\ and\ \citenamefont {Goldstein}}]{cisneros2011dynamics}%
  \BibitemOpen
  \bibfield  {author} {\bibinfo {author} {\bibfnamefont {L.~H.}\ \bibnamefont {Cisneros}}, \bibinfo {author} {\bibfnamefont {J.~O.}\ \bibnamefont {Kessler}}, \bibinfo {author} {\bibfnamefont {S.}~\bibnamefont {Ganguly}}, \ and\ \bibinfo {author} {\bibfnamefont {R.~E.}\ \bibnamefont {Goldstein}},\ }\href@noop {} {\bibfield  {journal} {\bibinfo  {journal} {Physical Review E—Statistical, Nonlinear, and Soft Matter Physics}\ }\textbf {\bibinfo {volume} {83}},\ \bibinfo {pages} {061907} (\bibinfo {year} {2011})}\BibitemShut {NoStop}%
\bibitem [{\citenamefont {Sampat}\ \emph {et~al.}(2022)\citenamefont {Sampat}, \citenamefont {Verma}, \citenamefont {Gupta},\ and\ \citenamefont {Mishra}}]{sampat2022ordering}%
  \BibitemOpen
  \bibfield  {author} {\bibinfo {author} {\bibfnamefont {P.~B.}\ \bibnamefont {Sampat}}, \bibinfo {author} {\bibfnamefont {A.}~\bibnamefont {Verma}}, \bibinfo {author} {\bibfnamefont {R.}~\bibnamefont {Gupta}}, \ and\ \bibinfo {author} {\bibfnamefont {S.}~\bibnamefont {Mishra}},\ }\href@noop {} {\bibfield  {journal} {\bibinfo  {journal} {Physical Review E}\ }\textbf {\bibinfo {volume} {106}},\ \bibinfo {pages} {054149} (\bibinfo {year} {2022})}\BibitemShut {NoStop}%
\bibitem [{\citenamefont {Valsakumar}\ \emph {et~al.}(1997)\citenamefont {Valsakumar}, \citenamefont {Satyanarayana},\ and\ \citenamefont {Sridhar}}]{valsakumar1997signature}%
  \BibitemOpen
  \bibfield  {author} {\bibinfo {author} {\bibfnamefont {M.}~\bibnamefont {Valsakumar}}, \bibinfo {author} {\bibfnamefont {S.}~\bibnamefont {Satyanarayana}}, \ and\ \bibinfo {author} {\bibfnamefont {V.}~\bibnamefont {Sridhar}},\ }\href@noop {} {\bibfield  {journal} {\bibinfo  {journal} {Pramana}\ }\textbf {\bibinfo {volume} {48}},\ \bibinfo {pages} {69} (\bibinfo {year} {1997})}\BibitemShut {NoStop}%
\bibitem [{\citenamefont {Maryshev}\ \emph {et~al.}(2019)\citenamefont {Maryshev}, \citenamefont {Goryachev}, \citenamefont {Marenduzzo},\ and\ \citenamefont {Morozov}}]{maryshev2019dry}%
  \BibitemOpen
  \bibfield  {author} {\bibinfo {author} {\bibfnamefont {I.}~\bibnamefont {Maryshev}}, \bibinfo {author} {\bibfnamefont {A.~B.}\ \bibnamefont {Goryachev}}, \bibinfo {author} {\bibfnamefont {D.}~\bibnamefont {Marenduzzo}}, \ and\ \bibinfo {author} {\bibfnamefont {A.}~\bibnamefont {Morozov}},\ }\href@noop {} {\bibfield  {journal} {\bibinfo  {journal} {Soft Matter}\ }\textbf {\bibinfo {volume} {15}},\ \bibinfo {pages} {6038} (\bibinfo {year} {2019})}\BibitemShut {NoStop}%
\bibitem [{\citenamefont {Strogatz}(2018)}]{strogatz2018nonlinear}%
  \BibitemOpen
  \bibfield  {author} {\bibinfo {author} {\bibfnamefont {S.~H.}\ \bibnamefont {Strogatz}},\ }\href@noop {} {\emph {\bibinfo {title} {Nonlinear dynamics and chaos: with applications to physics, biology, chemistry, and engineering}}}\ (\bibinfo  {publisher} {CRC press},\ \bibinfo {year} {2018})\BibitemShut {NoStop}%
\bibitem [{\citenamefont {Wolf}\ \emph {et~al.}(1985)\citenamefont {Wolf}, \citenamefont {Swift}, \citenamefont {Swinney},\ and\ \citenamefont {Vastano}}]{wolf1985determining}%
  \BibitemOpen
  \bibfield  {author} {\bibinfo {author} {\bibfnamefont {A.}~\bibnamefont {Wolf}}, \bibinfo {author} {\bibfnamefont {J.~B.}\ \bibnamefont {Swift}}, \bibinfo {author} {\bibfnamefont {H.~L.}\ \bibnamefont {Swinney}}, \ and\ \bibinfo {author} {\bibfnamefont {J.~A.}\ \bibnamefont {Vastano}},\ }\href@noop {} {\bibfield  {journal} {\bibinfo  {journal} {Physica D: nonlinear phenomena}\ }\textbf {\bibinfo {volume} {16}},\ \bibinfo {pages} {285} (\bibinfo {year} {1985})}\BibitemShut {NoStop}%
\bibitem [{\citenamefont {Kodba}\ \emph {et~al.}(2004)\citenamefont {Kodba}, \citenamefont {Perc},\ and\ \citenamefont {Marhl}}]{kodba2004detecting}%
  \BibitemOpen
  \bibfield  {author} {\bibinfo {author} {\bibfnamefont {S.}~\bibnamefont {Kodba}}, \bibinfo {author} {\bibfnamefont {M.}~\bibnamefont {Perc}}, \ and\ \bibinfo {author} {\bibfnamefont {M.}~\bibnamefont {Marhl}},\ }\href@noop {} {\bibfield  {journal} {\bibinfo  {journal} {European journal of physics}\ }\textbf {\bibinfo {volume} {26}},\ \bibinfo {pages} {205} (\bibinfo {year} {2004})}\BibitemShut {NoStop}%
\end{thebibliography}%
\end{document}